\documentclass[traditabstract]{aa}
\usepackage{natbib}
\usepackage{amssymb}
\usepackage{amsmath}
\usepackage{graphicx}
\usepackage{verbatim}
\usepackage{subfig}
\usepackage{epsfig}
\usepackage{tikz}
\usepackage{tikz-3dplot}
\usepackage{color}
\usepackage{multirow}
\usepackage{rotating}
\usepackage{textcomp}
\usepackage{booktabs}

\newcommand{\z}{[\ion{Fe}{}\!/\ion{H}{}\!]}
\newcommand{\teff}{$T_{\mathrm{eff}}$}
\newcommand{\taufh}{$\tau_{_{500\mathrm{nm}}}$}
\newcommand{\logg}{\mbox{log\,{\it g}}}
\newcommand{\dep}[1]{{\it b}$_{#1}$}

\newcommand{\fig}[1]{Fig.~\ref{#1}}

\newcommand{\sub}[1]{_{_{#1}}}

\newcommand{\Mgi}{\ion{Mg}{i}}
\newcommand{\Mgii}{\ion{Mg}{ii}}
\newcommand{\acorr}[1]{ \Delta \mathrm{A (#1) }}%_{_{\rm{NLTE-LTE}}}}

\begin{document}
% My data
\title{Mg line formation in late-type stellar atmospheres:\\ II. Calculations in a grid of 1D models\footnote{departure coefficients only available in electronic form at the CDS via anonymous ftp to {\tt cdsarc.u-strasbg.fr} (130.79.128.5) or via {\tt http://cdsweb.u-strasbg.fr/cgi-bin/qcat?J/A+A/} } }
\author{
  Y. Osorio \and  
  P. S. Barklem 
}
\institute{
Theoretical Astrophysics, Department of Physics and Astronomy, Uppsala University, Box 516, SE-751 20 Uppsala, Sweden}
\offprints{Yeisson Osorio}
\mail{yeisson.osorio@physics.uu.se}
\titlerunning{\ion{Mg}{} Line formation in late-type stellar atmospheres.}
\authorrunning{Osorio Y. \& Barklem P. S.}

% Doc

   \date{}
  \abstract
   {Mg is \emph{the} $\alpha$ element of choice for Galactic population and chemical evolution studies because it is easily detectable in all late-type stars. Such studies require precise elemental abundances, and thus departures from local thermodynamic equilibrium (LTE) need to be accounted for.} %leave it empty if necessary  
   {Our goal is to provide reliable departure coefficients and equivalent widths in non-LTE, and for reference in LTE, for diagnostic lines of Mg studied in late-type stars.  These can be used, for example, to correct LTE spectra and abundances.}
   {Using the model atom built and tested in the preceding paper in this series, we performed non-LTE radiative transfer calculations in a grid of 3\,945 stellar 1D atmospheric models. We used a sub-grid of 86 models to explore the propagation of errors in the recent atomic collision calculations to the radiative transfer results.}
   {We obtained departure coefficients for all the levels and equivalent widths (in LTE and non-LTE) for all the radiative transitions included in the `final' model atom presented in the preceding paper in this series. Here we present and describe our results and show some examples of applications of the data. The errors that result from uncertainties in the collisional data are investigated and tabulated.  The results for equivalent widths and departure coefficients are made freely available.}
  { Giants tend to have negative abundance corrections while dwarfs have positive, though small, corrections. Error analysis results show that uncertainties related to the atomic collision data are typically on the order of 0.01 dex or less, although for few stellar models in specific lines uncertainties can be as large as 0.03~dex.   As these errors are less than or on the same order as typical corrections, we expect that we can use these results to extract Mg abundances from high-quality spectra more reliably than from classical LTE analysis. Table \ref{fileweq} only available in electronic form at the CDS via anonymous ftp to {\tt cdsarc.u-strasbg.fr} (130.79.128.5) or via {\tt http://cdsweb.u-strasbg.fr/cgi-bin/qcat?J/A+A/} and the same data is accessible via the INSPECT project {\tt http://inspect.coolstars19.com}}
  % {Here is the abstract.}
   \keywords{NLTE --- line: formation --- stars: abundances}

   \maketitle

\section{Introduction}

Magnesium plays an important role in Galactic population and chemical evolution studies. It is released during type II supernovae occurring at the end of the life of massive stars, and so tracing the evolution of Mg is an important tool for our understanding of the evolution of the Galaxy.  Mg lines are observed in all late-type stars, even in very metal-poor stars \citep[e.g.][]{2002Natur.419..904C,2005Natur.434..871F} and in a wide spectral range from the UV to the IR. 

There have been a number of previous non local thermodynamic equilibrium (non-LTE) studies of Mg in different types of stars \citep[e.g.][]{1962ApJ...135..500A,2000ARep...44..530S,2001A&A...369.1009P,Langangen:2009jw,2013A&A...550A..28M}, with particular interest in the \Mgi\ emission lines observed in the Sun and some giant stars \citep{1988ApJ...330.1008M,1991ApJ...379L..79C,1992A&A...253..567C,1996ASPC..109..723U,2008A&A...486..985S}. The main conclusions from this collection of studies is the increase of non-LTE effects with decreasing metallicity (\z) and the demonstration that the IR Mg emission lines form because of non-LTE effects and have a photospheric origin. In the first paper of this series \cite[][hereafter Paper I]{MgPaperI}, we constructed a new Mg model atom for non-LTE studies in late-type stellar atmospheres; new quantum mechanical calculations were performed and implemented in the model atom. Comparison of the non-LTE line profiles with observed spectra and an investigation of the effect of different collisional processes in four stellar atmospheric models was also performed in Paper~I.

In this paper, we present a grid of non-LTE calculations for Mg in late-type stars, based on this model atom. Such large-scale calculations in grids of atmospheric models have been performed for a number of elements, e.g. lithium \citep[][using 72 and 392 atmospheric models respectively]{1994A&amp;A...288..860C,2009A&A...503..541L}, carbon \citep[176 models]{2006A&amp;A...458..899F}, sodium \citep[764 model atmospheres]{2011A&amp;A...528A.103L}, and iron \citep[$\sim$2900 models]{2012MNRAS.427...50L}. A grid for Mg has been calculated recently by \cite{2011MNRAS.418..863M} using $\sim$ 450 models for Mg lines with a model atom that neglects collisions with \ion{H}{}\ and uses electron collisional data from quantum mechanical calculations (when available), though the sources are not specified.

For this work, we use the MARCS\footnote{\url{http://marcs.astro.uu.se/}} grid of stellar atmospheres \citep{Gustafsson:2008df} to perform non-LTE calculations of Mg using the final model atom, called `F' in  Paper~I, and present non-LTE, and for reference LTE, equivalent widths.  From the equivalent widths, corresponding abundance corrections can be calculated for a set of abundances in each atmosphere. 

This paper is structured as follows.  In Sect.~\ref{sect:model} we briefly describe the model atom, the atmospheric models used, and the non-LTE spectrum calculations; the details are in  Paper~I. In Sect.~\ref{sect:results} we present the results and analyse the abundance corrections on the stellar grid (Sect.~\ref{sect:acorr}), study the propagation of uncertainties that result from uncertainties in the collisional data used (Sect.~\ref{sect:error}), and show examples of possible applications of our results (Sect.~\ref{sect:diffana}).  Finally in Sect.~\ref{sect:concl}, we present our conclusions.

\section{NLTE modelling}\label{sect:model}

Here we recap the most important aspects of the model atom, methods, and codes used for solving the non-LTE radiative transfer problem.  We then describe the application to the large-scale calculations performed in this work.

For solving the restricted non-LTE radiative transfer (RT) problem we used the {\tt MULTI} code \citep{MULTIuppsala,MULTIrev}, which uses the plane-parallel geometry setup and treats a given atomic species, here Mg, in the trace element approximation.   In this work, for the calculation of Mg non-LTE line formation, we use the  so called `final' model atom built and tested in  Paper I, Model F. This atom has 143 levels (108 for \Mgi\, 34 for \Mgii\ and the ground state of \ion{Mg}{iii}) and 1185 lines. Energy levels and $f$ values were taken mostly from the NIST database \citep{NIST}. Stark broadening data was taken mostly from  \cite{1996A&amp;AS..117..127D} and van der Waals broadening from the ABO theory \citep{ABO} and new calculations also presented in  Paper~I. New electron collisional data was calculated in  Paper~I and used in the F model atom, while for transitions not covered by the new calculations, we used the impact parameter (IP) method \citep{1962amp..conf..375S}. Hydrogen collisional data between low-lying states was taken from \cite{2012A&A...541A..80B} and the method of \cite{Kaulakys:1986tl} used for remaining transitions. Additionally, some variations of this model atom were created for testing the sensitivity of the results to the new collisional data from \cite{2012A&A...541A..80B} and that from  Paper~I (see Sect. \ref{sect:error}). For more details about the model atom see  Paper~I.

We performed non-LTE calculations for Mg in 3\,945 MARCS stellar atmospheric models with ranges of effective temperature \teff=$[2500, 8000]$~K, surface gravity \logg=$[-0.5, 5.0]$ (cm$/$s$^2$), and metallicity \z=$[-5.0, 1.0]$~dex. For each stellar model we used depth-independent micro-turbulent velocities of $\xi$=\,1.0 and 2.0~km/s and, additionally, $\xi$=\,3.0~km/s for the stellar models with \logg~$\leq3.5$~(cm$/$s$^2$).  Approximately 70\% of the MARCS atmospheric models used in this work are spherical model atmospheres, the remaining 30\% are plane-parallel models used for dwarfs, where the high surface gravity means the atmosphere is sufficiently thin for this approximation to be valid. 
The use of spherical atmospheric models in plane-parallel RT solvers was studied by \cite{2006A&A...452.1039H}. They recommend the use of spherical atmospheric models in plane-parallel RT solvers when spherical atmospheric models, but not spherical RT solvers, are available. However, they point out that for low \teff\ and \logg\ (cool giant) models, a plane-parallel atmosphere/RT~solver combination provides results that are closer to the spherical model atmosphere/RT~solver combination than the hybrid spherical model atmosphere/plane-parallel RT solver (see \cite{2006A&A...452.1039H} for details). 

For each stellar model, we calculated LTE and non-LTE equivalent widths ($w^*$ and $w$ respectively) for 23 abundance points between A(Mg)=2.6 and 8.6~dex\footnote{We use usual notation for element abundances A(X)=$\log_{10}\frac{n_X}{n_{\rm{H}}}$+12, where $n_x$ is the number density of species X.}. This allows abundance corrections to be deduced for a wide range of possible values of [Mg/Fe].  A subset of 86  stars was also used for studying how the uncertainties in the recent collisional calculations propagate to the final non-LTE equivalent widths and abundance corrections. 

In the calculations, we made use of the two options for starting procedures in {\tt MULTI}, which were found to lead to the most reliable convergence. For low gravity models LTE populations are used as a starting point. For high \logg\ stellar models, convergence was attained more efficiently with starting populations that correspond to the solution of the statistical equilibrium equations with zero radiation field.  

\section{Results}\label{sect:results}

We calculated departure coefficients \dep{i} of the 143 levels of the `F' model atom in  Paper~I. We also calculated equivalent widths in LTE and non-LTE for all the 1\,185 radiative transitions in the `F' model atom, but focus on a subset of 19 lines for which analysis is performed. The behaviour of Mg abundance corrections, $\acorr{Mg}$, defined as \[\acorr{Mg}$=A(Mg)$\sub{NLTE}-$A(Mg)$\sub{LTE},\] varies across the stellar grid, with magnesium abundance, and with the line under study. 

\subsection{Departure coefficients}\label{sect:dep}

For each stellar model and Mg abundance calculated, we obtained departure coefficients\footnote{Defined here as the ratio between \mbox{non-LTE} and LTE populations for level $i$, \dep{i}$=n_i/n_i^*$.} \dep{i} as a function of optical depth at 500~nm ($\tau\sub{500\,\rm{nm}}$) of the 143 levels used in the `F' model atom from  Paper~I.  The departure coefficients can be used in radiative transfer (RT) codes to include non-LTE effects in spectral fitting codes without the need to solve the non-LTE RT problem, e.g. the SME code \citep[][ and updates]{Valenti96} can use \dep{i}'s in this way (as demonstrated in Sec. 4.1 of  Paper I).   The behaviour of \dep{i} with A(Mg) is smooth, yet not simple,  especially for the low-lying excited levels of \Mgi; see for example \fig{fig:dep}, which shows how $b_i$ for the first excited level of \Mgi\ varies with Mg abundance and $\tau\sub{500\,\rm{nm}}$ for two stars differing only in metallicity.  Thus, results must be tabulated and to obtain reliable results in possible interpolation, the behaviour of \dep{i} with A(Mg) must be sufficiently sampled.  Each of the 3\,945 MARCS model atmospheres has 56 depth points (distributed between values of the Rosseland mean optical depth $\tau\sub{\rm{Ross}}=10^{-5}$ and $\tau\sub{\rm{Ross}}=100$, see \citeauthor{2008A&A...486..951G}, 2008 for details) for which we obtained Mg populations for each of the 143 levels in the model atom (together with other data), and we calculated 23 Mg abundance values.  The resulting departure coefficients are also available at CDS.

\def\angEl{30} % elevation angle
\def\angAz{300} % azimuth angle
       
\begin{figure*}[!ht]
\begin{tabular}{r@{\hspace{-0.15\textwidth}}l}
\hspace{-0.11\textwidth}\begin{tikzpicture}[>=latex,%
  % Create the xy-plane
  xzplane/.estyle={%
 cm={%
          cos(\angAz+90) , sin(\angAz+90)*sin(\angEl),%
          0              , cos(\angEl)            ,%
          (0,0)%
      }%
  },
  % Create the xy-plane
  xyplane/.estyle={%
      cm={%
          cos(\angAz)  , sin(\angAz)*sin(\angEl),%
          -sin(\angAz) , cos(\angAz)*sin(\angEl),%
          (0,0)%
      }%
  }]

\node[anchor=south west, inner sep=0] (image) at (0,0) { 
\subfloat{\includegraphics[width=0.7\textwidth]{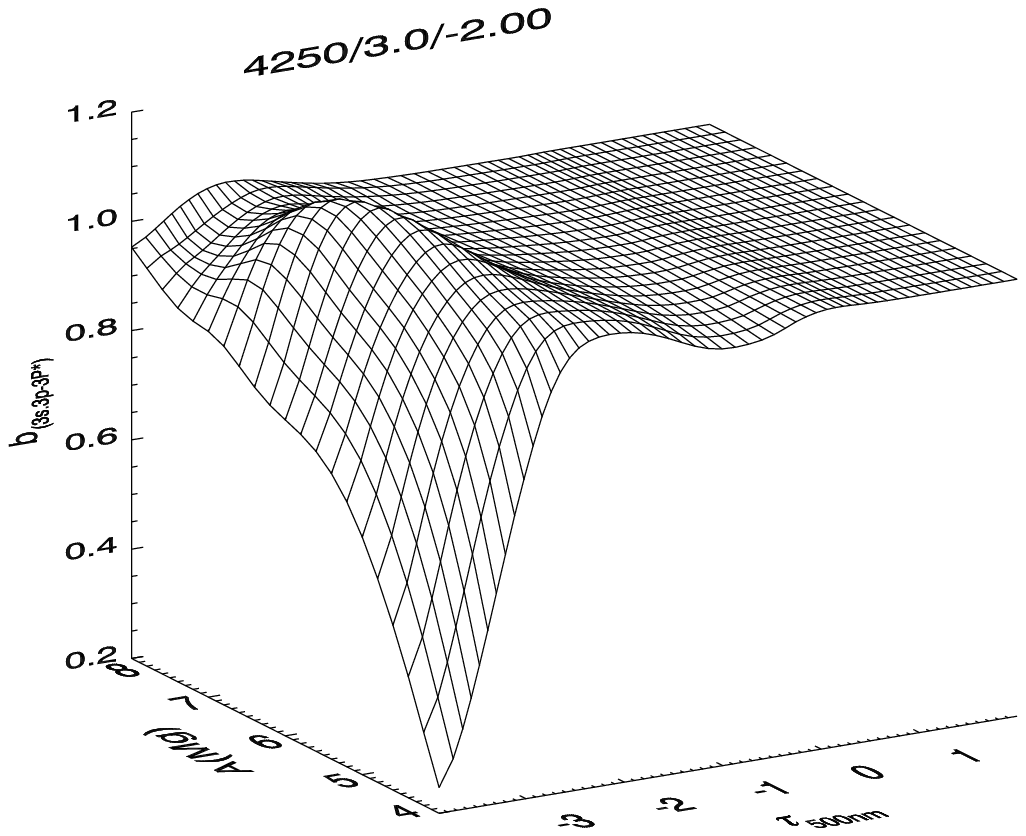}}
};
\draw  [fill=white,white] (0.11\textwidth,0.2\textwidth) rectangle (0.135\textwidth,0.3\textwidth);
\node[xzplane,rotate=90] at (0.12\textwidth,0.25\textwidth) {$b$ (3s.3p$-^3$P)};
\draw  [fill=white,white] (0.45\textwidth,0.04\textwidth) rectangle (0.55\textwidth,0.065\textwidth);
\node[font=\LARGE,xyplane,rotate=85] at (0.5\textwidth,0.06\textwidth) {$\tau_{500\rm{nm}}$};
\end{tikzpicture}
&
\hspace{-0.05\textwidth}\begin{tikzpicture}[>=latex,%
  % Create the xy-plane
  xzplane/.estyle={%
 cm={%
          cos(\angAz+90) , sin(\angAz+90)*sin(\angEl),%
          0              , cos(\angEl)            ,%
          (0,0)%
      }%
  },
  % Create the xy-plane
  xyplane/.estyle={%
      cm={%
          cos(\angAz)  , sin(\angAz)*sin(\angEl),%
          -sin(\angAz) , cos(\angAz)*sin(\angEl),%
          (0,0)%
      }%
  }]
\node[anchor=south west, inner sep=0] (image) at (0,0) { 
\subfloat{\includegraphics[width=0.7\textwidth]{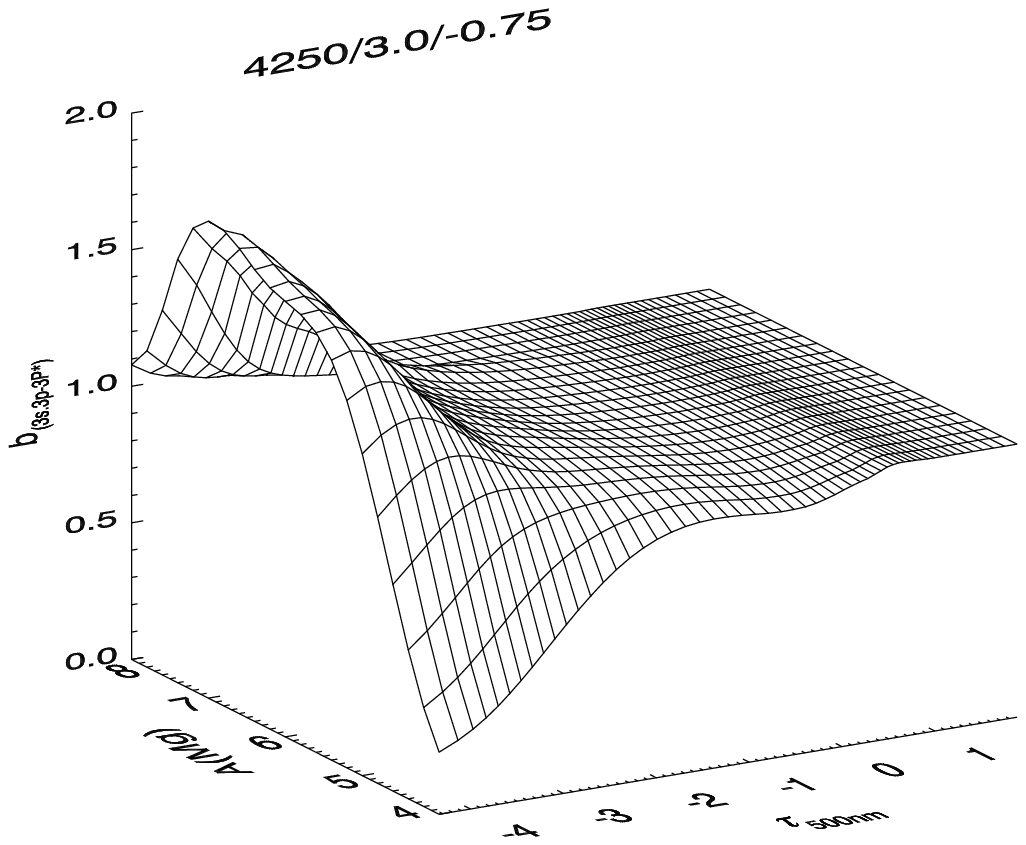}}
};
\draw  [fill=white,white] (0.11\textwidth,0.2\textwidth) rectangle (0.135\textwidth,0.3\textwidth);
\node[xzplane,rotate=90] at (0.12\textwidth,0.25\textwidth) {$b$ (3s.3p$-^3$P)};
\draw  [fill=white,white] (0.45\textwidth,0.04\textwidth) rectangle (0.55\textwidth,0.065\textwidth);
\node[font=\LARGE,xyplane,rotate=85] at (0.5\textwidth,0.06\textwidth) {$\tau_{500\rm{nm}}$};
\end{tikzpicture}
\end{tabular}
\caption{Departure coefficient \dep{}\ of the first excited level of \Mgi\ in stellar models with \teff/\logg/=4250/3.0 and \z=$-2.0$ (left) and $-0.75$ (right) as function of optical depth (at 500 nm) and Mg abundance. }\label{fig:dep}
\end{figure*}

\subsection{Equivalent widths and abundance corrections}\label{sect:acorr}

We present a set of non-LTE, and for reference LTE, equivalent widths for 19 diagnostic lines of Mg commonly observed in late-type stellar spectra. Abundances range from A(Mg)=2.6 to 8.6~dex sampled with 23 values for each of the 3\,945 1D stellar atmospheric models used from the MARCS grid.   Table \ref{fileweq} shows the format of the file available at CDS.  The first four columns show parameters of the stellar model \teff, \logg, \z, and micro-turbulent velocity $\xi$, the fifth column is the abundance of Mg, A(Mg), used, and the remaining columns report the predicted LTE equivalent widths $w^*$ and non-LTE equivalent widths $w$ for each line $\lambda$, in the dimensionless form $log_{10}(w/\lambda)$. 

% 3829.357            3832.302            3838.292            4167.271            4571.095            4702.990            5167.320            5172.683            5183.603            5528.403            5711.086            7691.548            8710.185            8712.686            8717.812            8736.012            8806.751            21059.727           21060.951

\begin{figure*}[!t]
\centering
\hspace{0.0\textwidth}\begin{tikzpicture}
\node[anchor=south west, inner sep=0] (image) at (0,0) {
\subfloat{\includegraphics[width=0.95\textwidth]{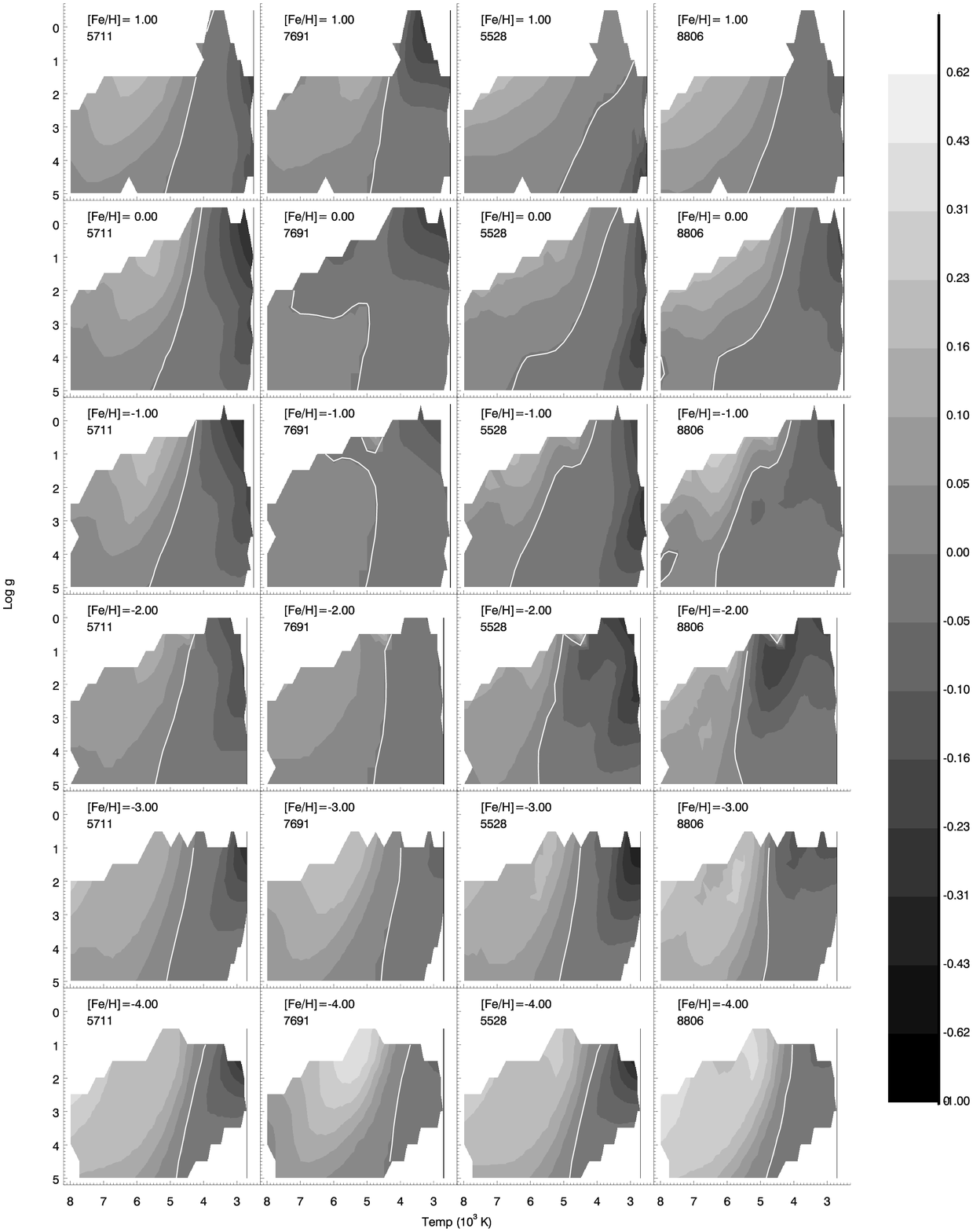}}
};
%\node at (0.345\textwidth,0.954\textwidth) {\scalebox{0.7}{${\color{white} +}$}};
%\node at (0.366\textwidth,0.954\textwidth) {\scalebox{0.7}{${\color{white} -}$}};

%\node at (0.335\textwidth,0.647\textwidth) {\scalebox{0.7}{${\color{white} +}$}};
%\node at (0.36\textwidth,0.647\textwidth) {\scalebox{0.7}{${\color{white} -}$}};

%\node at (0.36\textwidth,0.285\textwidth) {\scalebox{0.7}{${\color{white} +}$}};
%\node at (0.38\textwidth,0.285\textwidth) {\scalebox{0.7}{${\color{white} -}$}};
\draw [fill=white,white] (0.86\textwidth,0.115\textwidth) rectangle (0.89\textwidth,0.14\textwidth);
\node at (0.87\textwidth,0.125\textwidth) {\scalebox{0.6}{-1.00}};
\node[rotate=-90] at (0.9\textwidth,0.6\textwidth) {$\acorr{Mg}$};
%\node[anchor=south east] at (0.44\textwidth,-0.0065\textwidth) {
%\subfloat{\includegraphics[width=0.6\textwidth]{figures/panel_higz.eps}}
%};
%\draw [fill=white,white] (-0.15\textwidth,0.28\textwidth) rectangle (-0.142\textwidth,0.314\textwidth);
\end{tikzpicture}
\caption{Abundance corrections at [Mg/Fe]=0 for [Fe/H]$\ge0$ and at [Mg/Fe]=0.4 for [Fe/H]$<0$ in some of the atmospheric models of the stellar grid used in this work. Micro-turbulent velocity $\xi$=2.0~km/s in all the models in the plot. Each column shows corrections for a different \Mgi\ line. Note that the colour scaling is not linear. The white line shows the contour of zero abundance correction.}\label{fig:panel}
\end{figure*}

We selected 19 lines of \Mgi, some of them presented in Table 1 of  Paper~I\footnote{On request, we can provide similar data for any of the 1185 lines in the model atom described in  Paper~I.}, those lines are commonly used for Mg studies of late-type stars. For example the \Mgi\ b lines, which we included, are observed in a wide range of stellar parameters, even in very metal-poor stars. The inter-combination line at 4571~\AA and the 4703~\AA\  line, used commonly for Mg abundance determinations, are also included. Among the studied transitions there are the Mg lines covered by the GALAH (5711 and 7691~\AA) and Gaia-ESO (5528 and 8806~\AA) surveys. For the purposes of our discussion, we focus on these last four lines as representative of the typical behaviour of the abundance corrections $\acorr{Mg}$, defined here as the difference between the abundance in non-LTE minus the abundance in LTE corresponding to the same $w$ for the given Mg line. 
The abundance corrections derived from our equivalent width results for these lines are shown in \fig{fig:panel}.

\begin{table*}
\begin{center}
\caption{Segment of the file containing the LTE and non-LTE equivalent width results}\label{fileweq}
\scriptsize
\begin{verbatim}
* values in LTE and NLTE are log(w/lambda). lambda is in Angstroms
* Teff    log(g)    [Fe/H]    vturb    A(mg)        3832.302      ...      4571.095      ...      8806.751      ...      21060.951           
                                                 lte       nlte   ...   lte       nlte   ...   lte       nlte   ...   lte       nlte   
2500      0.00      0.50      1.00      2.60    -4.7924   -4.7737 ...  -5.2677   -5.3026 ...  -7.2368   -7.1734 ...  -8.6819   -8.6889
2500      0.00      0.50      1.00      3.00    -4.7067   -4.6879 ...  -5.0257   -5.0540 ...  -6.8477   -6.7835 ...  -8.2821   -8.2972
2500      0.00      0.50      1.00      3.60    -4.6042   -4.5880 ...  -4.8065   -4.8230 ...  -6.2937   -6.2245 ...  -7.6827   -7.7025
2500      0.00      0.50      1.00      4.00    -4.5328   -4.5165 ...  -4.7189   -4.7298 ...  -5.9634   -5.8844 ...  -7.2839   -7.3029
2500      0.00      0.50      1.00      4.60    -4.3910   -4.3701 ...  -4.6290   -4.6349 ...  -5.5557   -5.4554 ...  -6.6901   -6.7017
2500      0.00      0.50      1.00      5.00    -4.2618   -4.2372 ...  -4.5845   -4.5885 ...  -5.3450   -5.2310 ...  -6.3022   -6.3047
2500      0.00      0.50      1.00      5.30    -4.1466   -4.1208 ...  -4.5558   -4.5588 ...  -5.2140   -5.0926 ...  -6.0214   -6.0160
...       ...       ...       ...       ...       ...       ...   ...    ...       ...   ...    ...       ...   ...    ...       ... 
4500      4.50      0.00      2.00      7.60    -2.7603   -2.7606 ...  -4.1783   -4.1790 ...  -3.9022   -3.8962 ...  -4.3822   -4.3810
4500      4.50      0.00      2.00      7.80    -2.7069   -2.7072 ...  -4.1113   -4.1117 ...  -3.8295   -3.8242 ...  -4.2726   -4.2708
4500      4.50      0.00      2.00      8.00    -2.6650   -2.6652 ...  -4.0398   -4.0400 ...  -3.7624   -3.7577 ...  -4.1852   -4.1829
4500      4.50      0.00      2.00      8.10    -2.6484   -2.6486 ...  -4.0031   -4.0033 ...  -3.7314   -3.7270 ...  -4.1490   -4.1465
4500      4.50      0.00      2.00      8.30    -2.6233   -2.6235 ...  -3.9297   -3.9298 ...  -3.6749   -3.6709 ...  -4.0892   -4.0863
4500      4.50      0.00      2.00      8.50    -2.6070   -2.6071 ...  -3.8583   -3.8584 ...  -3.6263   -3.6228 ...  -4.0429   -4.0394
4500      4.50      0.00      2.00      8.60    -2.6013   -2.6014 ...  -3.8242   -3.8243 ...  -3.6050   -3.6017 ...  -4.0236   -4.0199
4500      4.50      0.25      1.00      2.60    -4.9143   -4.9958 ...  -7.3019   -7.4501 ...  -7.0945   -7.3881 ...  -9.1885   -8.9062
4500      4.50      0.25      1.00      3.00    -4.7183   -4.7723 ...  -6.9030   -7.0421 ...  -6.6981   -6.9274 ...  -8.7885   -8.5428
4500      4.50      0.25      1.00      3.60    -4.5119   -4.5408 ...  -6.3086   -6.4194 ...  -6.1153   -6.2646 ...  -8.1885   -8.0462
\end{verbatim}
\end{center}
\end{table*}

The physical conditions in the atmosphere and the interaction of Mg with electrons, hydrogen atoms and photons determines the behaviour of the Mg populations, spectral lines and thus abundance corrections, which can be described for different regions of the stellar parameter space (see \fig{fig:panel}). For cool giant atmospheric models, corrections tend to be negative and the most extreme values are around \mbox{$\z\sim-2$} for the 5528 and 8806~\AA\ lines, while the 7691~\AA\ line in cool giants has stronger corrections with  increasing metallicities. F and G dwarf models have for most lines small positive corrections at \z $\gtrsim -3$ that increase at lower metallicities, in particular for the 7691~\AA\ line. The higher density in dwarfs tends to thermalise the populations and therefore the corrections weaken towards higher \logg. The transition from negative to positive corrections is more dependent on \teff\ than on \logg\ or \z.  

The effect of micro-turbulence $\xi$ in the non-LTE abundance corrections is expected to be small given that obtaining abundance corrections is a differential procedure with respect to equivalent widths $w$. The strongest dependence of the obtained corrections with $\xi$ is of the order of 0.02 dex and occurs when the line is almost saturated. Weak lines are formed in the same region of the atmosphere so non-LTE corrections behave similar for  different values of $\xi$. For medium strength lines, not yet saturated in the line core, micro-turbulence affects the entire profile but the core and wings of the line are formed at different depths where different non-LTE mechanisms may dominate \citep{2010EAS....43..189M} causing the LTE and non-LTE profiles to behave slightly differently with respect to each other at different $\xi$. Once the line saturates, the wings start to be independent of $\xi$ and the non-LTE effects start to depend again only on the same depth in the atmosphere (where the core is formed). Abundance corrections become less dependent on $\xi$ with increasing abundance for saturated lines. Below we describe the behaviour of $\acorr{Mg}$ for different lines in the stellar parameter space \{\teff, \logg, \z\}.

 At \teff$\sim5000$~K the populations of \Mgi\ and \Mgii\ are comparable and departure coefficients tend to be close to unity ($b_i\sim1$).  \Mgi\ is therefore \emph{not} a minority species at lower \teff;  \Mgi population is larger than that of \Mgii\ in line formation regions. At \teff $\lesssim$~3000~K the \Mgi\ ground level, 3s\,$^1$S, and its first two excited states, 3p\,$^{3,1}P$, are thermalised at \taufh$\ge-1$. The lower state of the 5711 and 5528~\AA\ lines is 3p\,$^1$P, which is strongly collisionally coupled with the ground level of \Mgi. At \teff$\sim$4500~K, when the ground level of \Mgii\ starts to thermalise, the \Mgi\ low-lying levels are still close to LTE populations leading to almost LTE conditions in all the lines formed in atmospheres around these values of \teff. There is a less clear dependency of the behaviour of \dep{i}'s with \logg\ and metallicity, though at higher \logg\ the range in \teff\ where the atomic levels are close to LTE conditions broadens (see \fig{fig:panel}).

By checking the \dep{i} we can study the interplay of the different population and de-population mechanisms that produce the final line profiles and $w$'s. At \teff~$\gtrsim$~5000~K \Mgi\ starts to become a minority species and the ground level of \Mgii\ is thermalised leading to underpopulated \Mgi\ levels with respect to LTE due to over-ionisation. The 5528 \AA\ line has negative $\acorr{Mg}$ at low \teff, increasing towards the giant region of the parameter space and lower metallicities due to \dep{j}/\dep{i}$<1$ which become $>1$ at high \teff, with $j>i$ (the location of the transition temperature changes with metallicity and surface gravity). The 5711~\AA\ line has a similar behaviour though not as strong. Other Mg lines have different behaviour. The \Mgi\ inter-combination line at 4571~\AA\ has only positive $\acorr{Mg}$ that increase towards high metallicities, due to the UV over-ionisation depopulation of the \Mgi\ ground level in comparison with LTE (\dep{3s\,^1\rm{S}}$\lesssim1$). The \Mgi\ b lines show no significant $\acorr{Mg}$ at low \teff\ but hot dwarfs have negative corrections at $\z\sim-2$ and in the most extreme metal-poor atmospheres ($-5<$[Fe/H]$<-3$), corrections become positive at intermediate \teff. The IR lines at 2.1~$\mu$m show only positive corrections at high metallicities and strong positive corrections start to appear when the lines becomes too weak to have photon losses that compensate for \dep{j}/\dep{i}>1 in the line formation region.

\subsubsection{Comparison with earlier works}\label{sect:comp}

The previous non-LTE study of Mg that is most similar to ours, in terms of the atomic data used, is the work of \cite{2013A&A...550A..28M}.  
% in which $gf$ values were tested in A-type stars and 
In that work, non-LTE corrections for late-type stars using the recent calculations for H collisions presented in \cite{2012A&A...541A..80B} (which we also used) were compared with those obtained using the Drawin formula.   The main differences between our work and that study are, in addition to the use of different non-LTE codes, differences in electron collisional data, treatment of hydrogen collisional excitation rates for high-lying levels, and the use of MAFAGS-ODF \citep{1997A&amp;A...323..909F} model atmosphere including different opacities by \citeauthor{2013A&A...550A..28M}.   Comparison for some of the models and lines presented in Table 5 in \citeauthor{2013A&A...550A..28M} are in Table \ref{tab:mas}. In general the agreement is reasonable, considering the differences mentioned above.

\begin{table*}
\caption{Comparison of the Mg abundance corrections $\acorr{Mg}$ at [Mg/Fe]=0.4 dex obtained from us (Oso) with those from \citeauthor{2013A&A...550A..28M} (2013, Mas) for selected lines and model atmospheres.}\label{tab:mas}
\scalebox{1.0}{
%\begin{tabular}{l r r@{$\hspace{25pt}$}r r@{$\hspace{25pt}$}r r@{$\hspace{25pt}$}r r@{$\hspace{25pt}$}  r r@{$\hspace{25pt}$}  r r}
\begin{tabular}{l r r r r r r r r r r r r}\hline\hline\\[-8pt]
\multicolumn{1}{c}{Model} & \multicolumn{2}{c}{3829~\AA} & \multicolumn{2}{c}{5172~\AA} & \multicolumn{2}{c}{4571~\AA} & \multicolumn{2}{c}{4703~\AA} & \multicolumn{2}{c}{5528~\AA} & \multicolumn{2}{c}{5711~\AA} \\ \cmidrule(lr){2-3} \cmidrule(lr){4-5} \cmidrule(lr){6-7} \cmidrule(lr){8-9} \cmidrule(lr){10-11} \cmidrule(lr){12-13}
                        & \small{Oso} & \small{Mas} & \small{Oso} & \small{Mas} & \small{Oso} & \small{Mas} & \small{Oso} & \small{Mas} & \small{Oso} & \small{Mas} & \small{Oso}  & \small{Mas} \\\hline\\[-8pt]
6000/4.0/$-$1.0   &            0.02  & 0.05  &           0.00 	&     0.04 	&   0.03 	& 0.08 &       0.01 	&       0.01 	& $-$0.03 	& $-$0.03 	&    0.03 & 0.04 \\
6000/4.0/$-$2.0   &            0.01  & 0.04  &     $-$0.03 	&     0.02   	&   0.05 	& 0.08 &       0.03 	&       0.03 	& $-$0.01 	& $-$0.02 	&    0.03 & 0.03 \\
6000/4.0/$-$3.0   &            0.05  & 0.08  &          0.01  	&      0.05  	&   0.12 	&          &       0.11 	&       0.07 	&      0.10 	&       0.07 	&    0.10 &  \\
5000/2.0/$-$1.0   &            0.04  & 0.06  &          0.03  	&      0.06  	&   0.20 	& 0.13 &       0.00 	& $-$0.07	& $-$0.07 	& $-$0.16 	&    0.02 & $-$0.05 \\
5000/2.0/$-$2.0   &            0.01  & 0.08  &     $-$0.03 	&      0.03  	&   0.13 	& 0.16 & $-$0.05 	& $-$0.02	& $-$0.13	& $-$0.12 	&    0.02 & 0.03 \\
5000/2.0/$-$3.0   &      $-$0.02  & 0.06  &     $-$0.12 	& $-$0.06	&   0.23 	& 0.29 &       0.10 	&       0.18	&       0.06	&       0.10 	&    0.07 &    \\
%
%
%6000/4.0/$-$1.0   &            0.019  & 0.05  &       0.003 & 0.04   &   0.031 & 0.08 &    0.005 & 0.01  &   $-$0.029 &  $-$0.03  &    0.031 & 0.04 \\
%6000/4.0/$-$2.0   &            0.013  & 0.04  &     $-$0.026 & 0.02   &   0.048 & 0.08 &    0.028 & 0.03  & $-$0.011 & $-$0.02   &    0.034 & 0.03 \\
%6000/4.0/$-$3.0   &            0.053  & 0.08  &      0.008  & 0.05   &   0.119 &          &    0.106 & 0.07  &     0.100 & 0.07    &    0.096 &  \\
%5000/2.0/$-$1.0   &            0.040  & 0.06  &      0.030  & 0.06   &   0.195 & 0.13 &   $-$0.002 & $-$0.07 &   $-$0.066 & $-$0.16  &     0.022 & $-$0.05 \\
%5000/2.0/$-$2.0   &            0.013  & 0.08  &     $-$0.034  & 0.03   &   0.129 & 0.16 & $-$0.051 & $-$0.02 &  $-$0.134  & $-$0.12 &    0.018 & 0.03 \\
%5000/2.0/$-$3.0   &      $-$0.016  & 0.06  &     $-$0.117 & $-$0.06 &  0.231 & 0.29 &  0.095  & 0.18 &      0.059  & 0.10 &    0.066 &    \\
\hline
\end{tabular}
}
\end{table*}

The study of \cite{2011MNRAS.418..863M} uses, like this work, the MULTI code and MARCS model atmospheres.   However, hydrogen collisions are not treated in their modelling.  Table \ref{tab:mer} in this work is an adaptation of their Table 2, comparing with some of our results.  The differences vary with spectral line and stellar parameters, but in general, our corrections are smaller (i.e. our results are closer to LTE).  Note that for the \z$=-2$ models the corrections suggested by \citeauthor{2011MNRAS.418..863M} are $>0.15$~dex while the corrections found by us are significantly smaller ($\sim0.02$~dex). 
 
\begin{table}
\caption{Comparison of the equivalent widths $w$ ($^*$ means LTE) and abundance corrections $\acorr{Mg}$ obtained by us (Oso) with those obtained by \citeauthor{2011MNRAS.418..863M} (2011, Mer) using the 8736~\AA\ line.}\label{tab:mer}
\scalebox{0.75}{
\begin{tabular}{l@{\hspace{5pt}}l@{\hspace{5pt}}l@{\hspace{5pt}}c r r r r r r} \hline\hline\\[-8pt]
         &             &          &        &     \multicolumn{6}{c}{Mg I 8736.012~\AA}       \\        
 {\tiny \teff (K)} &  {\tiny \logg}  &   {\tiny \z}  &    {\tiny [Mg/Fe]}  &   \multicolumn{2}{c}{{\tiny $w$[m\AA]}} & \multicolumn{2}{c}{{\tiny $w/w^*$}}    & \multicolumn{2}{c}{{\tiny $\acorr{Mg}$}} \\    
               &             &        &                   &   {\tiny Oso}  &  {\tiny Mer}&   {\tiny Oso}  &  {\tiny Mer}&   {\tiny Oso}  &  {\tiny Mer}\\\hline\\[-8pt]
 5000   &     1    &      +0.50   &   0.0   &  218  &  225  &     1.148   &   1.179  &   $-$0.27   & \\
    & &                        +0.25   &   0.0   &  183 &  205  &   1.164  &     1.226     &  $-$0.25   & \\
    & &                        +0.00   &   0.0     & 154 & 177   &    1.177 &    1.268     &   $-$0.23    & \\
    & &                        $-$0.25  &    0.1     &  135 & 158  &  1.176  &      1.299 & $-$0.21   &   \\              
     & &                       $-$0.50   &   0.2    &   118  & 135 &     1.171 &     1.302 &  $-$0.18  &  $-$0.11    \\    
      & &                      $-$0.75  &    0.3     &  102 & 111  &    1.170   &   1.272  &  $-$0.15   & $-$0.10     \\
       &  &                    $-$1.00  &    0.4    &     86  &  89   &     1.164 &     1.210 & $-$0.13   &  $-$0.08       \\
          &       &            $-$1.50  &    0.4    &    40  &   33  &   1.109     &    0.927 &  $-$0.06   &  +0.03      \\
           &      &            $-$2.00  &    0.4   &     14  &    9  &      0.985  &     0.667 & +0.01   &  +0.18       \\ 
           &        2   &     +0.50  &    0.0    &    217  & 211  &    1.077      &1.170     &      $-$0.11       &  \\
                        &   &  +0.25  &    0.0    &    180 & 192 &       1.088     &1.196  &      $-$0.12        &      \\
    &   &                      +0.00  &    0.0    &       148 & 167  &        1.094       &1.214 &   $-$0.12   &  \\
               &   &           $-$0.25  &    0.1    &    130 & 151 &     1.090     &1.230 &       $-$0.10   &  \\
   &   &                       $-$0.50  &    0.2   &     112 & 131 &     1.082    &1.236  &  $-$0.08  &$-$0.09    \\
              &   &            $-$0.75   &   0.3   &      97  & 110 &        1.080    &1.218  &  $-$0.07 &  $-$0.09     \\
                         &   & $-$1.00  &    0.4   &      82  & 88   &     1.073      &1.170  & $-$0.06  & $-$0.07      \\
       &   &                   $-$1.50  &    0.4  &       39  &  34   &      1.031     &0.919   &  $-$0.02  & +0.04      \\
                  &   &        $-$2.00  &    0.4   &      14  &  10   &      0.962       & 0.683  &  +0.02  & +0.17      \\\hline
\end{tabular}
}
\end{table}

\subsection{Sensitivity to new collisional data}\label{sect:error}

Collisional processes are a major source of uncertainty in non-LTE radiative transfer calculations. In this section we study the sensitivity  of the abundance corrections on the  atomic data calculated in  Paper~I for electron collisions and the calculations for hydrogen collisional data presented in \cite{2012JPhCS.397a2053G}, \cite{2012PhRvA..85c2704B} and \cite{2012A&A...541A..80B}. 

In order to check the sensitivity of our results on the new collisional data we built model atoms with 0.5 and 2.0 times the collisional rates calculated for excitation due to collisions with~\ion{H}{}\ (CH) and charge exchange with~\ion{H}{}\ (CH0) from \cite{2012JPhCS.397a2053G}, \cite{2012PhRvA..85c2704B} and \cite{2012A&A...541A..80B}. These multiplication factors represent an estimate of the typical error of those collisional rates, at least for the transitions with the largest rates.  Note, errors in transitions with very small rates may well be higher; however, we expect these rates to be less important in the modelling. The same factors (0.5 and 2.0) were used to study the sensitivity of the rates due to electron collisions (CE) calculated in  Paper~I,  although the error in the CE collisional rates is expected to be less for transitions between the lowest lying states, typically $\sim$20\% (<0.1 dex).
These new model atoms were used in non-LTE calculations on a sub-grid of 86 stellar models with  3500$\le$\teff[K]$\le$7500,    1$\le$\logg\,[cm/s$^{-2}$]$\le$4 and $-4\le$\z$\le$0.5 dex.

In general, we note that the interplay of the \Mgi\ level populations and different excitation/ionisation processes together with the transition from \Mgii\ to \Mgi\  as minority species, leads to different effects of the collisional processes on the final non-LTE populations. For example, in  Paper~I we pointed out that in some cases, adding collisional interactions to the calculations \emph{does not} necessarily lead to thermalisation.  

At very low \teff\ ($\lesssim$3500~K), the ground state of \Mgi\ (3s$^2\,^1$S) and the first two excited states (3p\,$^{3,1}$P) are thermalised so changes in the collisional data do not make their populations to depart from LTE in the line formation regions. The 3p\,$^1$P state is the lower level of the 5711 and the 5528~\AA\ lines, so abundance corrections of these lines come from the departure coefficients \dep{i} of the upper levels (5s\,$^1$S for the 5711~\AA\ and 4d\,$^1$D for the 5528 \AA\ lines).  \dep{5\rm{s}^1\rm{S}} and \dep{4\rm{d}^1\rm{D}} are less than unity in the line formation region, which strengthens the non-LTE lines when compared with LTE making the abundance corrections negative. The upper levels of these lines are sensitive mostly to CH, and so are the spectral lines involving them. CH0 does not affect the level populations in these atmospheres.  The 8806~\AA\ line has 3p\,$^1$P as its lower level and 3d\,$^1$D as its upper level. The 3d\,$^1$D level is particularly sensitive to CH in the atmospheric region where the core of the 8806~\AA\ line is formed leading to a CH-sensitive non-LTE equivalent width $w$.

\begin{figure*}[t!]
%\centering
\hspace{0.04\textwidth}\begin{tikzpicture}
\node[anchor=south east, inner sep=0] (image) at (0,0) {
\subfloat{\includegraphics[width=0.9\textwidth]{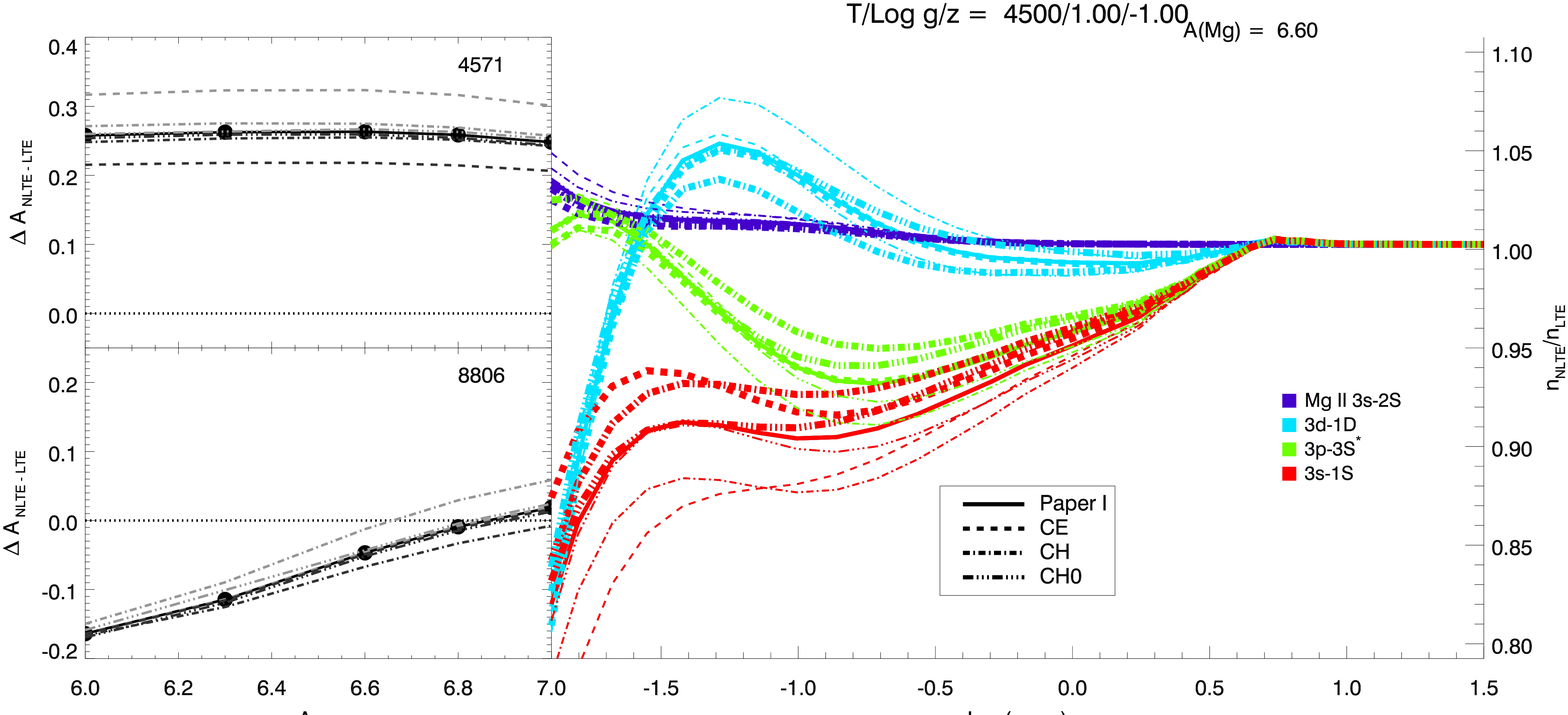}}
};
\draw [fill=white,white] (0.5\textwidth,-0.01\textwidth) rectangle (0.6\textwidth,0.0);
%\node[anchor=south east] at (0.44\textwidth,-0.0065\textwidth) {
%\subfloat{\includegraphics[width=0.6\textwidth]{figures/panel_higz.eps}}
%};
%\draw [fill=white,white] (-0.15\textwidth,0.28\textwidth) rectangle (-0.142\textwidth,0.314\textwidth);
\end{tikzpicture}
\caption{Sensitivity to collisional data in stellar model \teff/\logg/\z = 4500/1.0/-1.0.\\
{\bf Left}. Abundance corrections $\acorr{Mg}$ for the 4571~\AA\ (upper) and 8806~\AA\ (lower) lines. The solid lines and dots are $\acorr{Mg}$ calculated with the model atom F described in  Paper I. Broken lines are calculations done with modified collisional rates C$\to$C$\times$2.0 (black) and  C$\to$C$\times$0.5 (grey)  from the CE (dash), CH (dot-dash) or CH0 (three-dot-dash) rates used in the F model atom and derived from recent calculations (see text). \\
{\bf Right}. Departure coefficients at A(Mg)=6.6 of the levels involved in the 8806~\AA\ (3d\,$^1$D - 3p\,$^1$P) and the 4571(3p\,$^3$P - 3s\,$^1$S) transitions, together with the ground level of \Mgii, as function of optical depth. 3p\,$^1$P has a very similar behaviour to 3p\,$^3$P so it is not shown in the figure to aid visualisation. Solid  and broken lines have the same description as in the left figures; thick broken lines are for C$\times$2.0 and thin broken lines for C$\times$0.5 rate coefficients.}\label{fig:error}
\end{figure*}

At higher \teff\ ($\sim$4000 K), populations become sensitive to CH0 and CE. In these stars the \Mgi\ level populations are almost thermalised in the line formation region \mbox{(\taufh$\sim-0.5$)} though slightly underpopulated with respect to LTE; thermalisation is stronger at higher surface gravities. The CH0 rate coefficients of the levels around 4s\,$^1$S make ionising transitions of these levels collisionally dominated (C$_{ij}$ > R$_{ij}$) even at \taufh=$-3$ creating a strong coupling with the ground level of the \Mgii\ reservoir such that an increase in the collisional rates, leads to an increase in the populations of \Mgi\ levels.  Abundance corrections become sensitive to CE towards higher \teff, especially for the inter-combination line 4571~\AA. An increase of the CE rates, makes the population of  3s$^2\,^1$S increase, while for all the other levels the effect is the opposite making the upper levels of the lines involving 3s$^2\,^1$S become closer to this level and therefore reducing the non-LTE corrections. 
Sensitivity to CE increases with metallicity due to the fact that $n_{e}$ increases, and so does the electron collisional rates CE. [Mg/Fe] close to the solar values leads to larger absolute magnesium abundance at higher stellar metallicities leading to stronger lines whose cores form in higher layers in the atmosphere, where the populations become sensitive to CE. The lines with broad wings tend to be sensitive also to CH. 

For \teff~$\gtrsim$~5000~K the ground state of \Mgii\ is thermalised and \Mgi\ becomes a minority species. Changes in CH0 produce the biggest change in the departure coefficients of \Mgi\ levels. However, this does not necessarily mean that the abundance corrections are more sensitive to CH0 than to CH or CE. For example the 8806~\AA\ line has very little sensitivity to CH0 in atmospheres at these temperatures but the two levels involved in this transition (3p\,$^1$P and 3d\,$^1$D) are more sensitive to CH0 than to CH or CE in these atmospheres. CH0 affects their populations in such a way that the relative populations between those levels are not altered i.e. CH0 do not change the line source function. This leads to similar line profiles and therefore weak sensitivity of $\acorr{Mg}$ to CH0. CH and CE on the other hand affect the \Mgi\ levels in different ways: the first 4 low-lying \Mgi\ levels have larger populations when those rates are doubled and other levels lower their populations. This changes the relative populations of 3p\,$^1$P with respect to 3d\,$^1$D making the 8806~\AA\ line source function more sensitivity to CH and CE. The 8806~\AA\ line opacity on the other hand has similar sensitive to all collisional processes given that the departure coefficient for lower level of this line, \dep{3p\,^1P}, increases when any of these collision rates increase.     

Fig.~\ref{fig:error} shows an example where some of the features discussed above can be seen. By increasing CE rate coefficients, the ground level of \Mgi\ also increases in population while the \Mgi\ excited levels become slightly less populated. Decreasing CE has the opposite effect and the sensitivity of the abundance corrections on CE for the 4571~\AA\ line is quite significant while the 8806~\AA\ line has little sensitivity to the CE rates. When collisional excitation with CH is modified to CH$\times$2.0,
the population of the 3d\,$^1$D level is reduced (blue, thick, dot-dashed line in the right plot of \fig{fig:error}) compared with the original version of the `F' model atom (blue, solid line). The opposite happens with the 3p\,$^{1,3}$P levels (green lines). The total effect of CH$\to$CH~$\times$~2.0 on the 8806~\AA\ line (black, dot-dashed line in lower left corner of Fig.~\ref{fig:error}) is more negative abundance corrections than those obtained with the original CH rates.
Changes in the CH0 rate coefficients are more uniform among the \Mgi\ levels, using CH0~$\times$~2.0 leads to an increase of the populations of all the \Mgi\ levels and CH0~$\times$~0.5 leads to lower \Mgi\ populations (three-dot dashed lines in the right plot). This "uniform shift" of \dep{i} with increasing CH0 observed in \fig{fig:error} leads to less sensitivity of the non-LTE line profiles to charge exchange collisions for this particular stellar model. As mentioned before, different stars show different effects on different levels so the sensitivity of the non-LTE results to a given collisional process varies across the stellar parameter space. 

Table \ref{tab:error} shows the uncertainties associated with the collisional data calculated in  Paper I for CE and the data presented in \cite{2012PhRvA..85c2704B, 2012A&A...541A..80B} for CH and CH0. The uncertainties in $w$ associated with these collisional data are $<1\%$ (0.005~dex) for the vast majority of stars and lines. However, for a few stellar models, usually metal-poor, in specific lines and collisional processes, the difference in $w$ due to uncertainties in the collisional data can be as large as $7\%$ (0.03~dex).   One exception is change in the 4571~\AA\ line due to CE in the model \teff/\logg/\z=5500/4.0/-4.0, where the $w$ can be 17\% (0.07~dex) smaller if the CE collisional rates are lowered by 50\%.   Recall, however, that errors in CE collisional rates are expected to be somewhat smaller, $\sim 20$\%.  The sign in the uncertainty values is related to the direction of change in $w$ relative to the change in the collisional rates. For example, if the +CH error (i.e, make CH $\to$ CH~$\times$~2.0) has a negative value then increasing CH decreases $w$. Likewise,  if the $-$CH error (CH $\to$ CH~$\times$~0.5) is negative then decreasing CH increases $w$. These error propagation results reflect the high non-linearity of the non-LTE calculations. For a given stellar atmosphere, a line can have uncertainties in $w$ correlated to a given collisional process and anti-correlated to another, e.g. the star \teff/\logg/\z = 5500/4.0/+0.0 in the 8712~\AA\ line has CE \& CH anti-correlated errors and at the same time CH0 correlated errors. Moreover, in some cases, e.g. CH for the 5172~\AA\ line in the star 4500/3.0/$-$4.0, increasing and decreasing CH rates, both lead to smaller abundance corrections\footnote{By checking the line profiles of the 5172~\AA\ line, the CH$\times$2 predicted line has wider wings and a weaker core than the CH$\times$0.5 predicted line, leading to similar equivalent widths.}.

% !TEX root = Mg_paperII.tex
\begin{sidewaystable*}
\caption{ Non-LTE equivalent widths in the form log($w/\lambda$) for some Mg lines at [Mg/Fe]=0 and changes (in log($w$)) due to the use of C$\times$2.0 ($+$) or C$\times$0.5 ($-$) where C are the rates from the recent calculations on electron collisional excitation (CE) from Paper I, hydrogen collisional excitation (CH) and charge transfer with hydrogen (CH0) from \cite{2012A&A...541A..80B}. }\label{tab:error}
\scalebox{0.65}[0.58]{
\begin{tabular}{l c r r r c r r r c r r r c r r r c r r r c r r r}\hline\hline
\multicolumn{1}{r}{$\lambda$(\AA)\,\,$\to$}   & \multicolumn{4}{c}{ 4\,167}  & \multicolumn{4}{c}{ 4\,571}  & \multicolumn{4}{c}{ 5\,172}  & \multicolumn{4}{c}{ 5\,528}  & \multicolumn{4}{c}{ 8\,712}  & \multicolumn{4}{c}{21\,060}  \\ 

\multirow{2}{*}{ \teff/ \logg/\z } & \multirow{2}{*}{  log($w/\lambda$)} & \scalebox{0.8}{ $+$CE   } & \scalebox{0.8}{   $+$CH } & \scalebox{0.8}{    $+$CH0 } &  \multirow{2}{*}{  log($w/\lambda$)} & \scalebox{0.8}{ $+$CE   } & \scalebox{0.8}{   $+$CH } & \scalebox{0.8}{    $+$CH0 } &  \multirow{2}{*}{  log($w/\lambda$)} & \scalebox{0.8}{ $+$CE   } & \scalebox{0.8}{   $+$CH } & \scalebox{0.8}{    $+$CH0 } &  \multirow{2}{*}{  log($w/\lambda$)} & \scalebox{0.8}{ $+$CE   } & \scalebox{0.8}{   $+$CH } & \scalebox{0.8}{    $+$CH0 } &  \multirow{2}{*}{  log($w/\lambda$)} & \scalebox{0.8}{ $+$CE   } & \scalebox{0.8}{   $+$CH } & \scalebox{0.8}{    $+$CH0 } &  \multirow{2}{*}{  log($w/\lambda$)} & \scalebox{0.8}{ $+$CE   } & \scalebox{0.8}{   $+$CH } & \scalebox{0.8}{    $+$CH0 } \\ 
   &             & \scalebox{0.8}{ $-$CE   } & \scalebox{0.8}{   $-$CH } & \scalebox{0.8}{    $-$CH0 } &             & \scalebox{0.8}{ $-$CE   } & \scalebox{0.8}{   $-$CH } & \scalebox{0.8}{    $-$CH0 } &             & \scalebox{0.8}{ $-$CE   } & \scalebox{0.8}{   $-$CH } & \scalebox{0.8}{    $-$CH0 } &             & \scalebox{0.8}{ $-$CE   } & \scalebox{0.8}{   $-$CH } & \scalebox{0.8}{    $-$CH0 } &             & \scalebox{0.8}{ $-$CE   } & \scalebox{0.8}{   $-$CH } & \scalebox{0.8}{    $-$CH0 } &             & \scalebox{0.8}{ $-$CE   } & \scalebox{0.8}{   $-$CH } & \scalebox{0.8}{    $-$CH0 } \\
\hline

\end{tabular}
}
\end{sidewaystable*}

\subsection{Differential effects}\label{sect:diffana}

Applications of abundance studies are usually interested in using or comparing abundances from stars of different types, i.e. their relative abundances.  For example, it is of great interest to trace abundance changes with stellar evolutionary status among populations, e.g. globular clusters, to understand stellar processes.  Another example, is the study of differences in elemental abundances in various parts of the Galaxy (thin and thick disks and halo), and to identify new components.  The detection of new substructure will require high precision, and in order to identify stars of different types to a common population (sometimes called chemical tagging) or have the best statistics we must be able to accurately compare abundances from stars of different types \citep[e.g.][]{2013A&amp;A...553A..94L}.  

\begin{figure*}[t]
\scalebox{0.98}{\begin{tabular}{@{\hspace{-0.02\textwidth}}c@{\hspace{-0.045\textwidth}}c@{\hspace{-0.045\textwidth}}c}
\vspace{-0.05\textwidth}\begin{tikzpicture}
\node[anchor=south east, inner sep=0] (image) at (0,0) {
\subfloat{\includegraphics[width=0.38\textwidth]{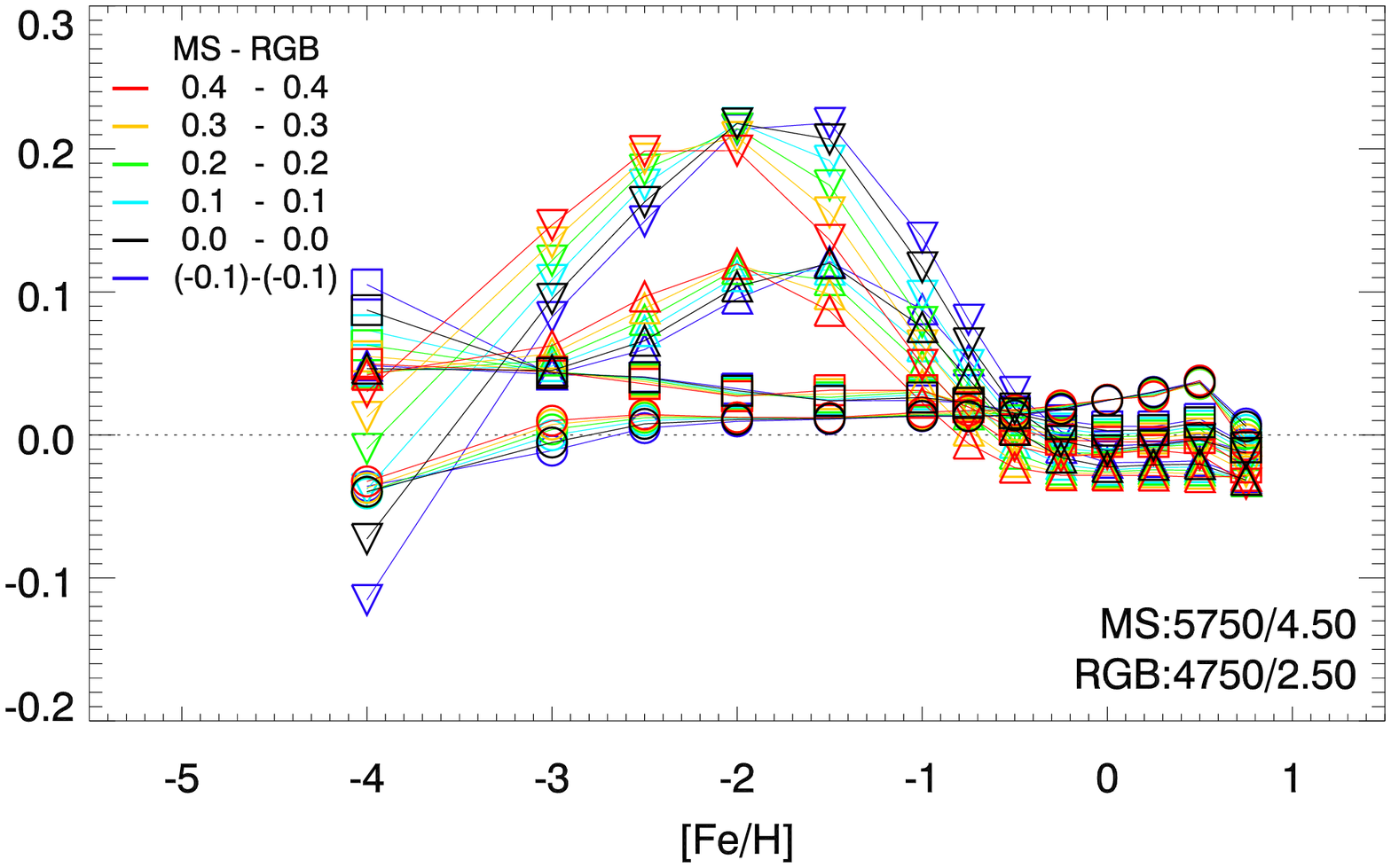}}
};
\node [rotate=90] at (-0.345\textwidth,0.12\textwidth) {\scalebox{0.6}{$\Delta\rm{[Mg/Fe]}_{{\rm{NLTE}}}$ (MS-RGB)}};
\end{tikzpicture}
& 
\begin{tikzpicture}
\node[anchor=south east, inner sep=0] (image) at (0,0) {
\subfloat{\includegraphics[width=0.38\textwidth]{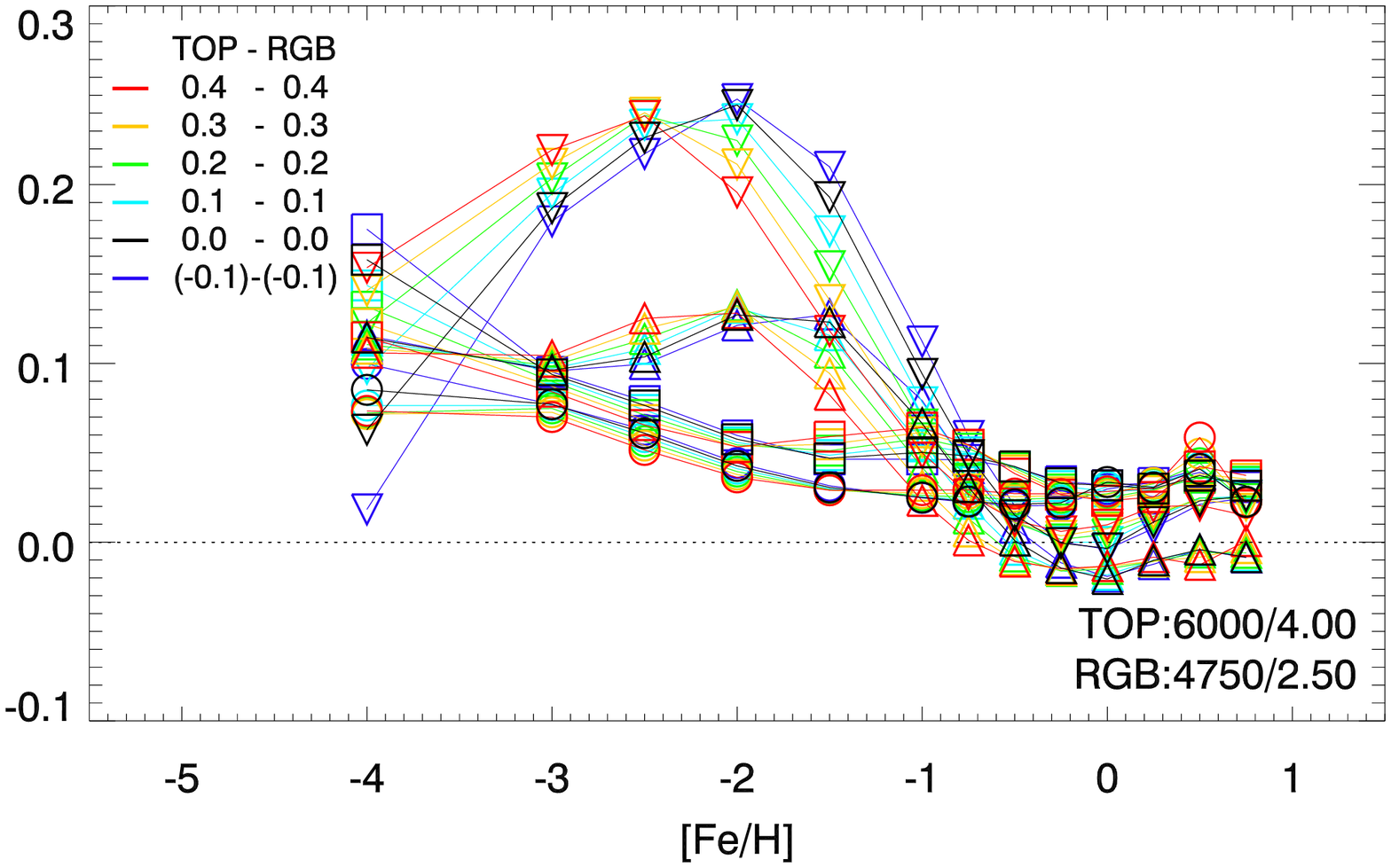}}
};
\node [rotate=90] at (-0.345\textwidth,0.12\textwidth) {\scalebox{0.6}{$\Delta\rm{[Mg/Fe]}_{{\rm{NLTE}}}$ (TOP-RGB)}};
\end{tikzpicture}
&  
\begin{tikzpicture}
\node[anchor=south east, inner sep=0] (image) at (0,0) {
\subfloat{\includegraphics[width=0.38\textwidth]{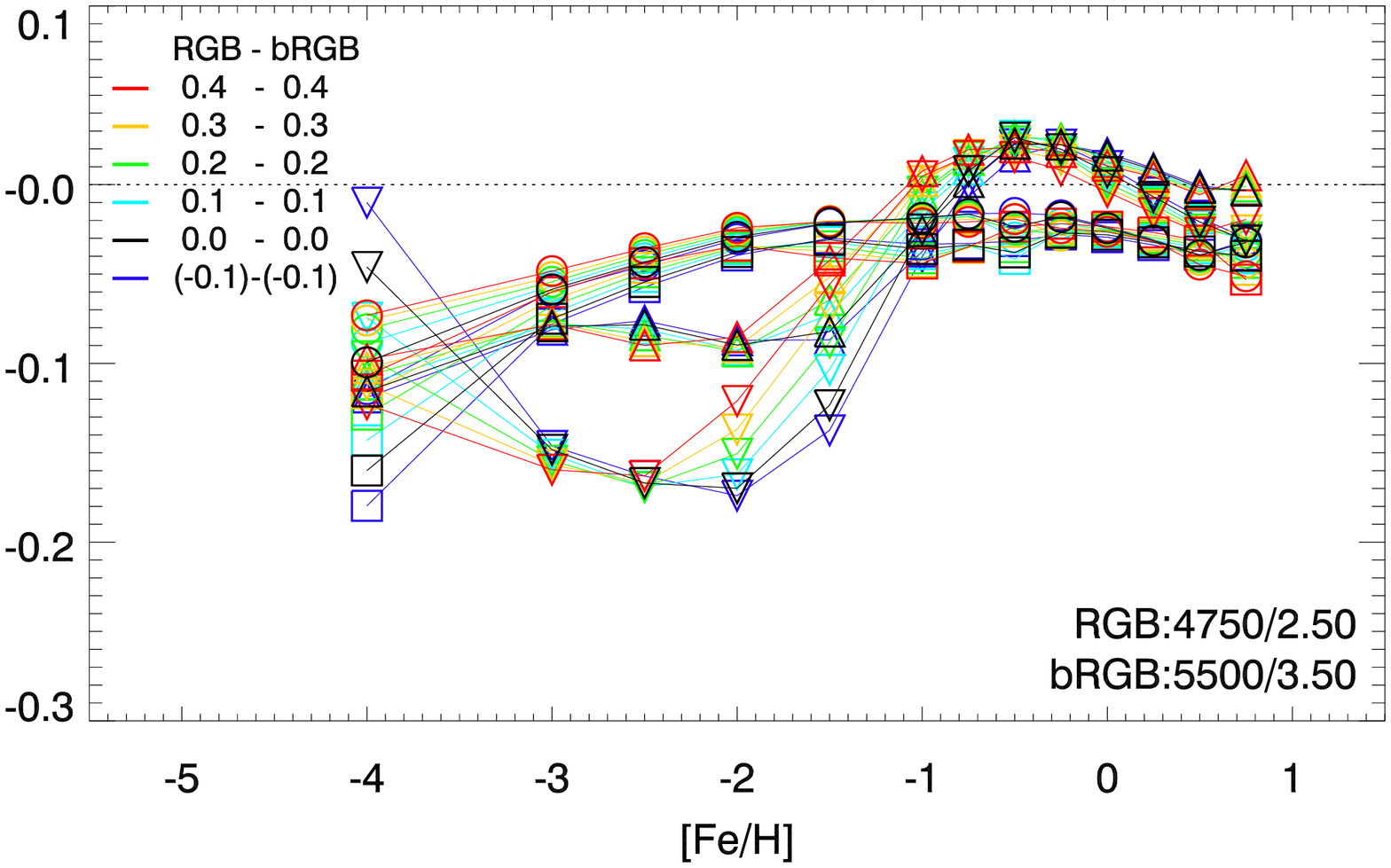}}
};
\node [rotate=90] at (-0.345\textwidth,0.12\textwidth) {\scalebox{0.6}{$\Delta\rm{[Mg/Fe]}_{{\rm{NLTE}}}$ (RGB-bRGB)}};
\end{tikzpicture}
\\
\vspace{-0.05\textwidth}\begin{tikzpicture}
\node[anchor=south east, inner sep=0] (image) at (0,0) {
\subfloat{\includegraphics[width=0.38\textwidth]{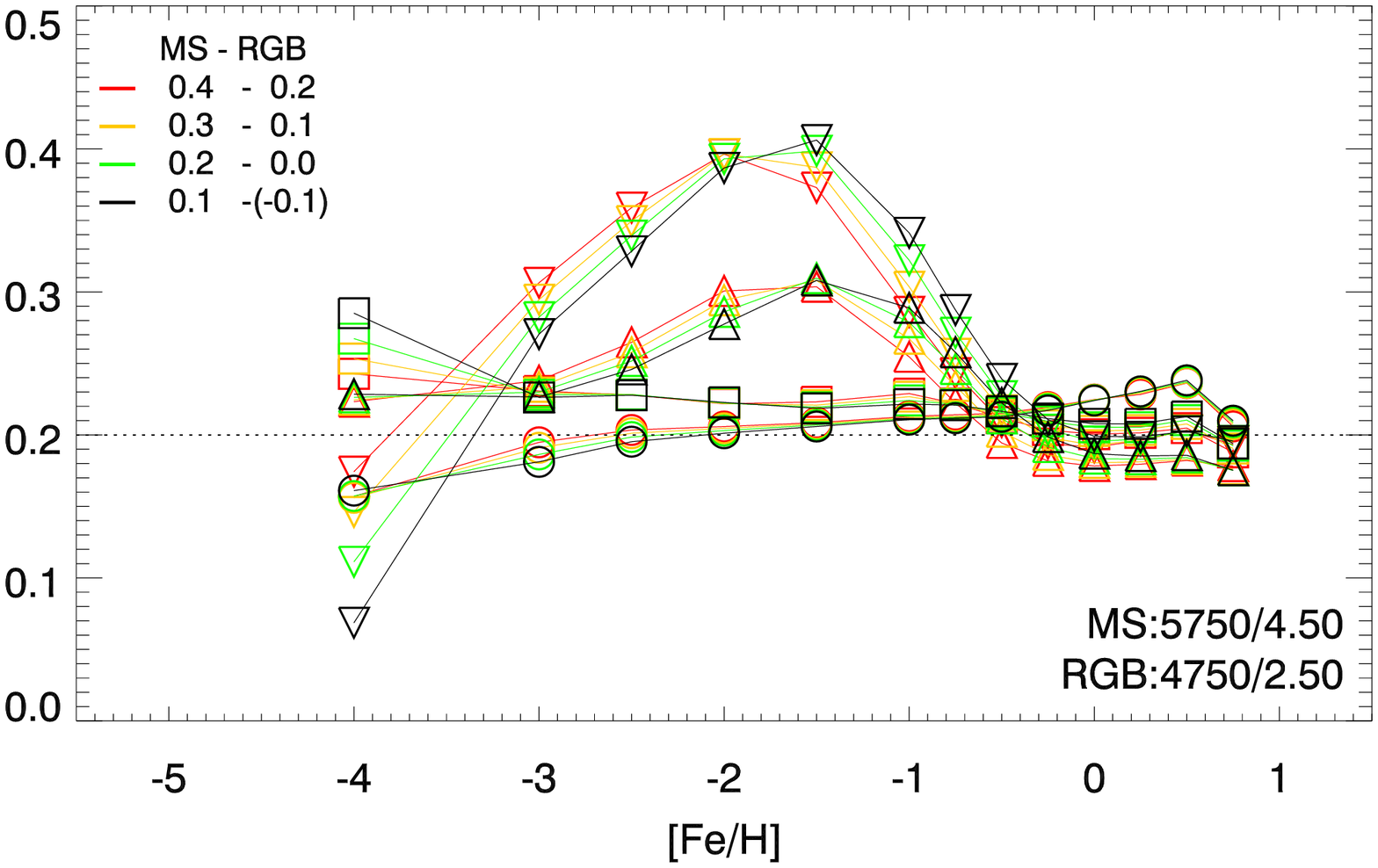}}
%\subfloat{\includegraphics[width=0.38\textwidth]{figures/MS-RGB.eps}}
};
\node [rotate=90] at (-0.345\textwidth,0.12\textwidth) {\scalebox{0.5}{$\Delta\rm{[Mg/Fe]}_{{\rm{NLTE}}}$ (MS-RGB)}};
\end{tikzpicture}
& 
\begin{tikzpicture}
\node[anchor=south east, inner sep=0] (image) at (0,0) {
\subfloat{\includegraphics[width=0.38\textwidth]{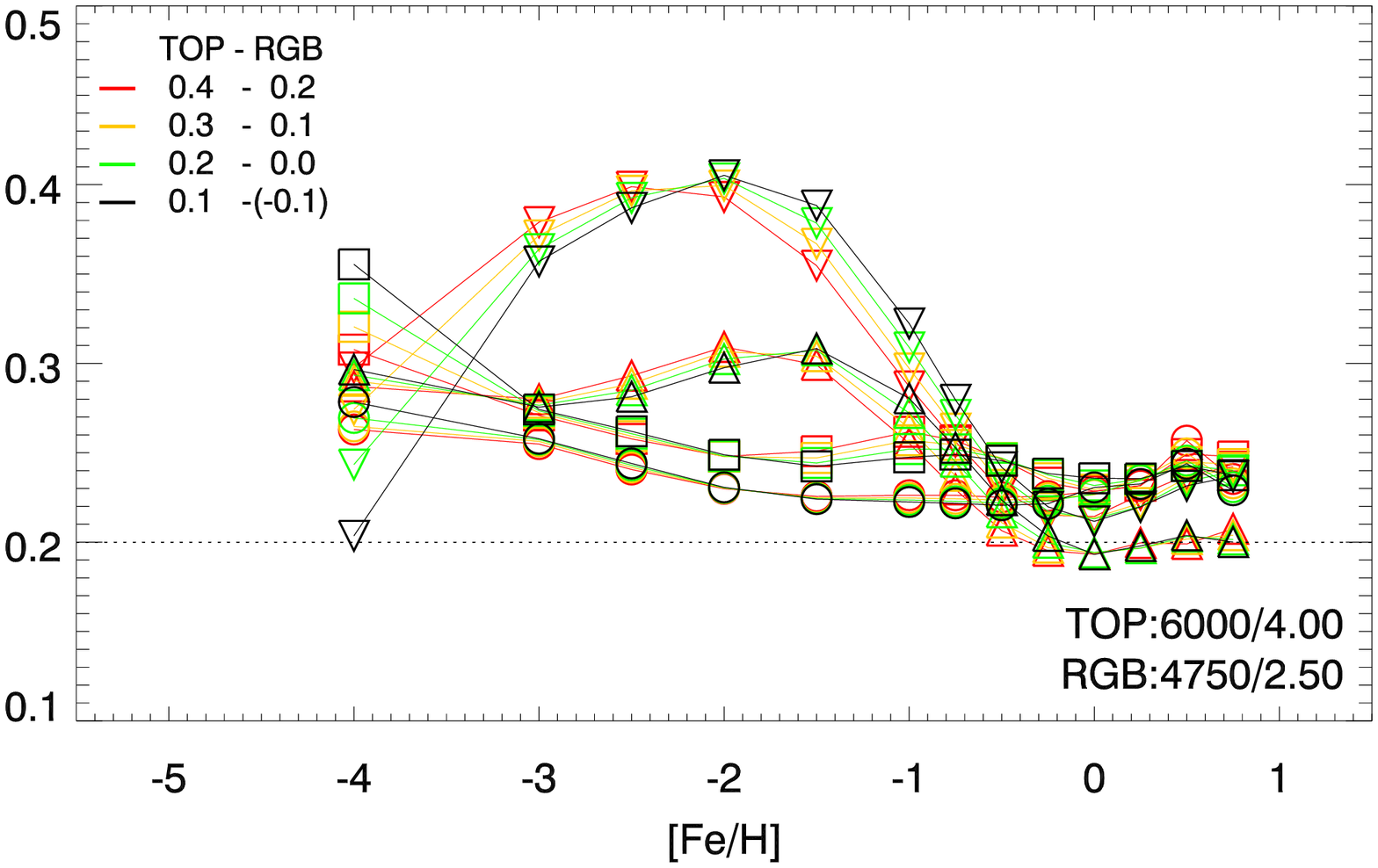}}
%\subfloat{\includegraphics[width=0.38\textwidth]{figures/TOP-bRGB.eps}}
};
\node [rotate=90] at (-0.345\textwidth,0.12\textwidth) {\scalebox{0.5}{$\Delta\rm{[Mg/Fe]}_{{\rm{NLTE}}}$ (TOP-RGB)}};
\node [left] at (-0.05\textwidth,0.194\textwidth) {\scalebox{0.6}{$\bigtriangleup$} \scalebox{0.6}{5528 \AA}};
\node [left] at (-0.05\textwidth,0.18\textwidth) {\scalebox{0.6}{$\Box$} \scalebox{0.6}{5711 \AA}};
\node [left] at (-0.05\textwidth,0.165\textwidth) {\scalebox{0.5}{$\bigcirc$} \scalebox{0.6}{7691 \AA}};
\node [left] at (-0.05\textwidth,0.15\textwidth) {\scalebox{0.6}{$\bigtriangledown$} \scalebox{0.6}{8806 \AA}};
\end{tikzpicture}
&  
\begin{tikzpicture}
\node[anchor=south east, inner sep=0] (image) at (0,0) {
\subfloat{\includegraphics[width=0.38\textwidth]{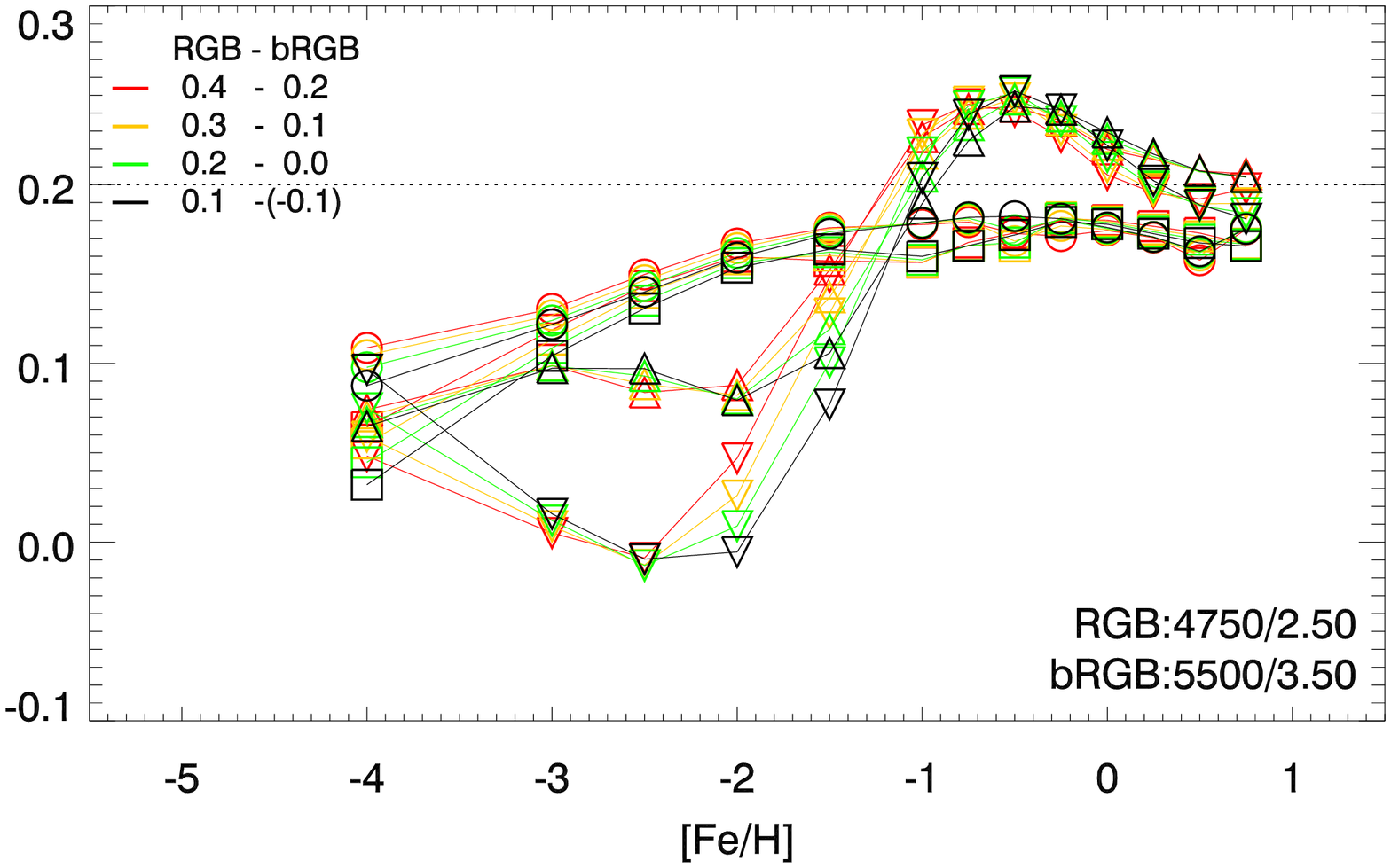}}
%\subfloat{\includegraphics[width=0.38\textwidth]{figures/bRGB-SGB.eps}}
};
\node [rotate=90] at (-0.345\textwidth,0.12\textwidth) {\scalebox{0.5}{$\Delta\rm{[Mg/Fe]}_{{\rm{NLTE}}}$ (RGB-bRGB)}};
\end{tikzpicture}
\\
\begin{tikzpicture}
\node[anchor=south east, inner sep=0] (image) at (0,0) {
\subfloat{\includegraphics[width=0.38\textwidth]{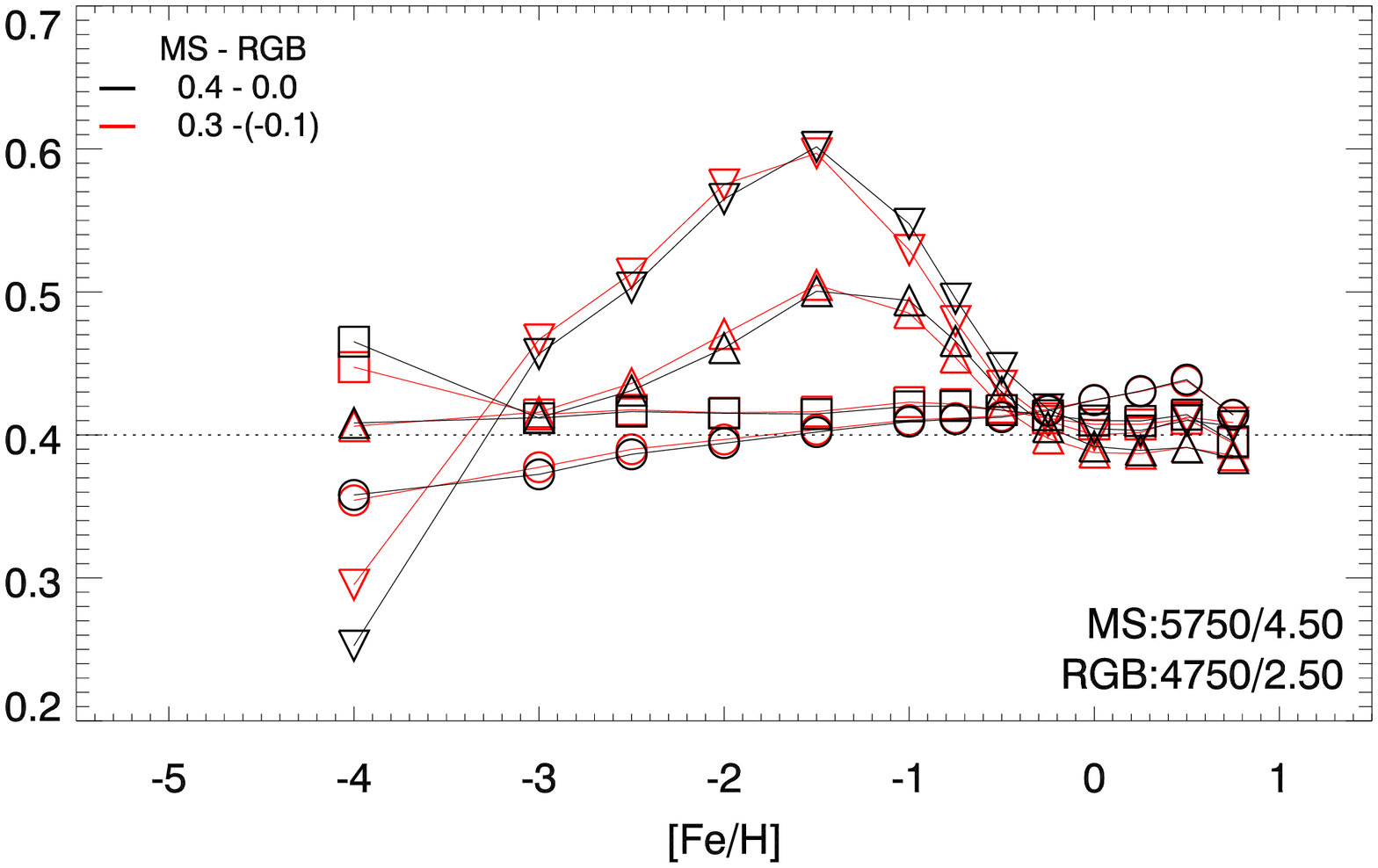}}
};
\node [rotate=90] at (-0.345\textwidth,0.12\textwidth) {\scalebox{0.5}{$\Delta\rm{[Mg/Fe]}_{{\rm{NLTE}}}$ (MS-RGB)}};
\end{tikzpicture}
& 
\begin{tikzpicture}
\node[anchor=south east, inner sep=0] (image) at (0,0) {
\subfloat{\includegraphics[width=0.38\textwidth]{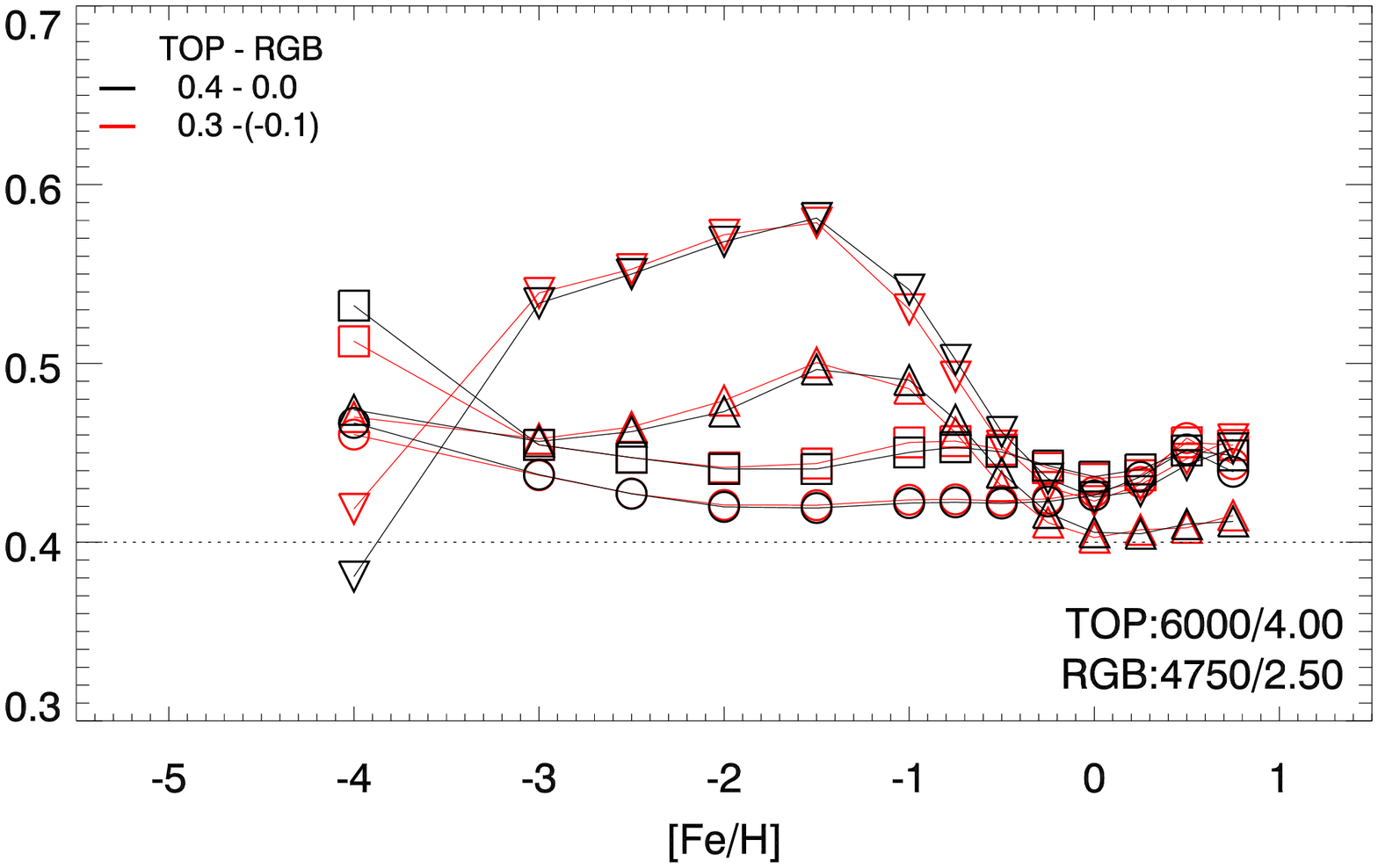}}
};
\node [rotate=90] at (-0.345\textwidth,0.12\textwidth) {\scalebox{0.5}{$\Delta\rm{[Mg/Fe]}_{{\rm{NLTE}}}$ (TOP-RGB)}};
\end{tikzpicture}
&  
\begin{tikzpicture}
\node[anchor=south east, inner sep=0] (image) at (0,0) {
\subfloat{\includegraphics[width=0.38\textwidth]{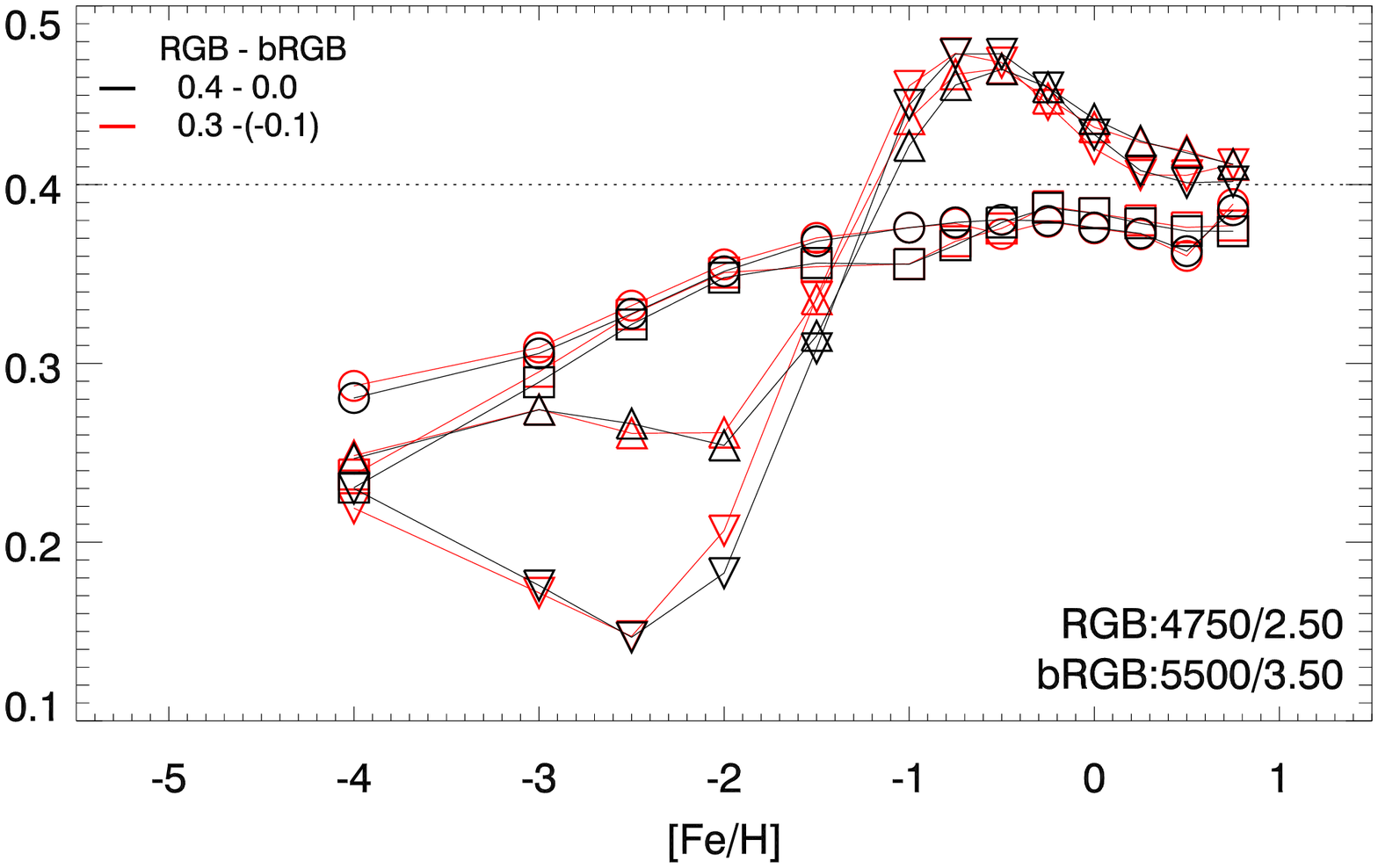}}
};
\node [rotate=90] at (-0.345\textwidth,0.12\textwidth) {\scalebox{0.5}{$\Delta\rm{[Mg/Fe]}_{{\rm{NLTE}}}$ (RGB-bRGB)}};
\end{tikzpicture}
\end{tabular}
}
\caption{Difference in non-LTE Mg abundance $\Delta$[Mg/Fe]$\sub{\rm{NLTE}}$ versus stellar metallicity for a fixed LTE difference $\Delta$[Mg/Fe]$\sub{\rm{LTE}}$ (=0, 0.2 and 0.4 in the first, second and third row respectively), between two stars with \teff/\logg\  parameters shown at the bottom right of each plot. The LTE difference $\Delta$[Mg/Fe]$\sub{\rm{LTE}}$ is represented by the dotted line. Colours indicate the LTE abundance [Mg/Fe]$_{_{\rm{LTE}}}$ for each star shown at the top-left of each plot. Four different \Mgi\ lines were used: 5711($\Box$), 7691($\bigcirc$), 5528 ($\bigtriangleup$) and 8806~($\bigtriangledown$)~\AA. Examples of individual stars at various evolutionary stages were selected: main sequence (MS), turn off point (TOP), red giant branch (RGB) and base of the RGB (bRGB). }\label{fig:diff}
\end{figure*}

An interesting question is how non-LTE affects the differences in abundances between stars with different stellar parameters, compared with LTE modelling.
In \fig{fig:diff}, each panel shows the difference in non-LTE abundance \[\Delta[\rm{Mg/Fe}]\sub{\rm{NLTE}}=[\rm{Mg/Fe}]\sub{\rm{A}_{NLTE}}-[\rm{Mg/Fe}]\sub{\rm{B}_{NLTE}}\] for two stars A and B with the same metallicity and difference in Mg LTE abundance $\Delta[\rm{Mg/Fe}]\sub{\rm{LTE}}$ (represented by the dotted line). We call  $\Delta[\rm{Mg/Fe}]\sub{\rm{NLTE}}$ the \emph{differential abundance correction}.  The first row in \fig{fig:diff} shows the difference in  $\Delta[\rm{Mg/Fe}]\sub{\rm{NLTE}}$ when the two stars have the same [Mg/Fe]$\sub{\rm{LTE}}$ abundance; different values of [Mg/Fe]$\sub{\rm{LTE}}$ are represented in different colours. Examples of individual stars at various evolutionary stages were selected: main sequence (MS), turn off point (TOP), red giant branch (RGB) and base of the RGB (bRGB). The most noticeable feature is the increase of non-LTE differential abundance correction with decreasing metallicity. Red giant stars have their largest abundance corrections for the 8806~\AA\ line around \z$=-2$, and as a consequence differences are stronger at these values of \z. For example if a metal-poor RGB star with \teff/\logg/\z=4750/2.5/-2.0 shows [Mg/Fe]$\sub{\rm{LTE}}$=0.1~dex and a bRGB star with \teff/\logg/\z=5500/3.5/-2.0 shows [Mg/Fe]$\sub{\rm{LTE}}=-0.1$~dex, the LTE abundance difference is 0.2~dex (black lines in the right, middle plot in \fig{fig:diff}) but the non-LTE abundance  difference obtained using the 8806~\AA\ ($\bigtriangledown$) line is 0.0~dex and using the 5528 \AA\ ($\bigtriangleup$) line is 0.1~dex.

\begin{figure*}
\scalebox{0.95}{\begin{tabular}{@{\hspace{-0.04\textwidth}}c@{\hspace{-0.07\textwidth}}c@{\hspace{-0.07\textwidth}}c}
\vspace{-0.06\textwidth}\begin{tikzpicture}
\node[anchor=south east, inner sep=0] (image) at (0,0) {
\subfloat{\includegraphics[width=0.41\textwidth]{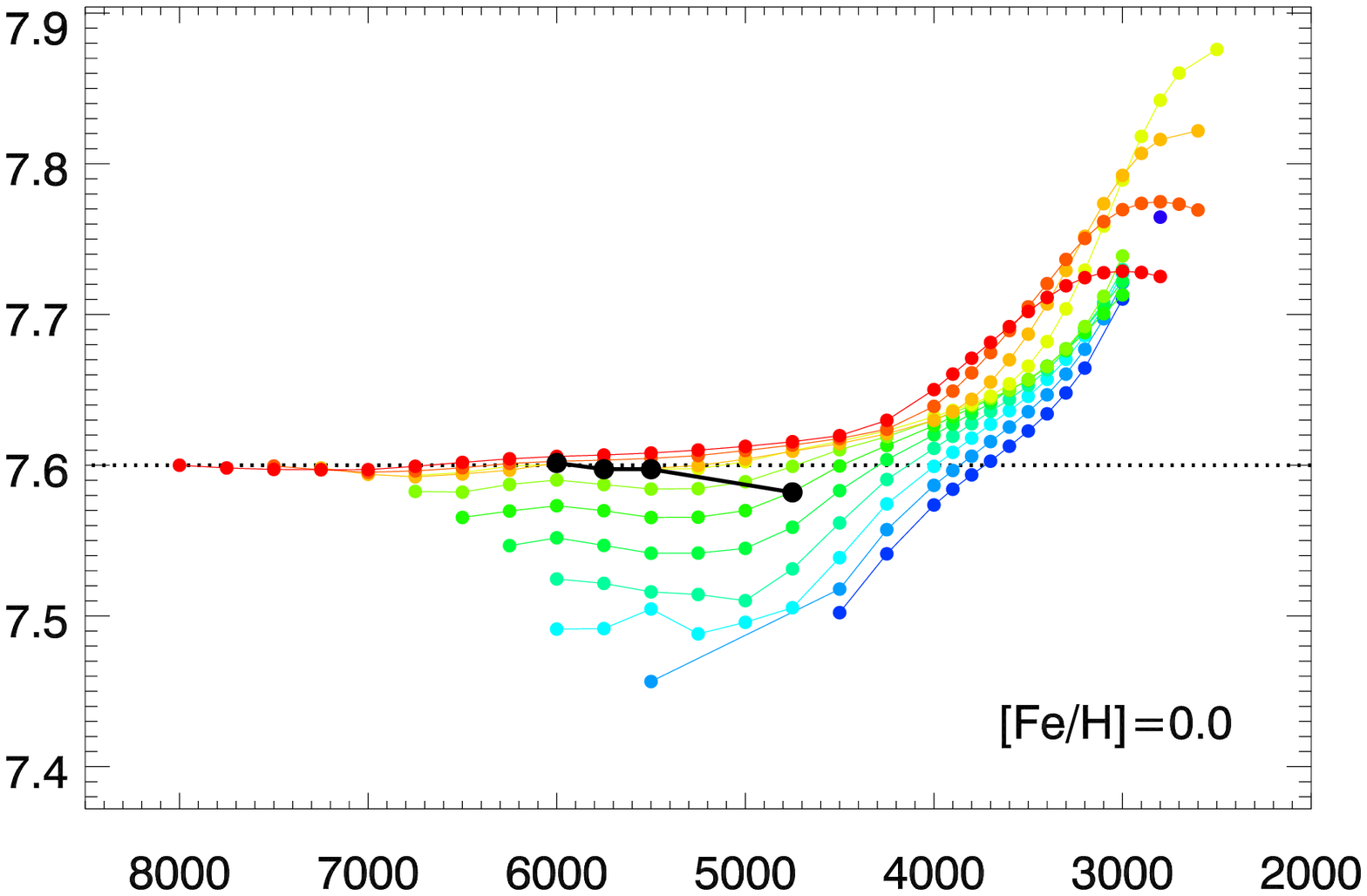}}
};
\node at (-0.29\textwidth,0.23\textwidth) {\scalebox{0.8}{5528 \AA}}; 
\node [rotate=90] at (-0.36\textwidth,0.15\textwidth) {\scalebox{0.7}{A(Mg)$_{_{\rm{LTE}}}$}};
\end{tikzpicture}
&
\begin{tikzpicture}
\node[anchor=south east, inner sep=0] (image) at (0,0) {
\subfloat{\includegraphics[width=0.41\textwidth]{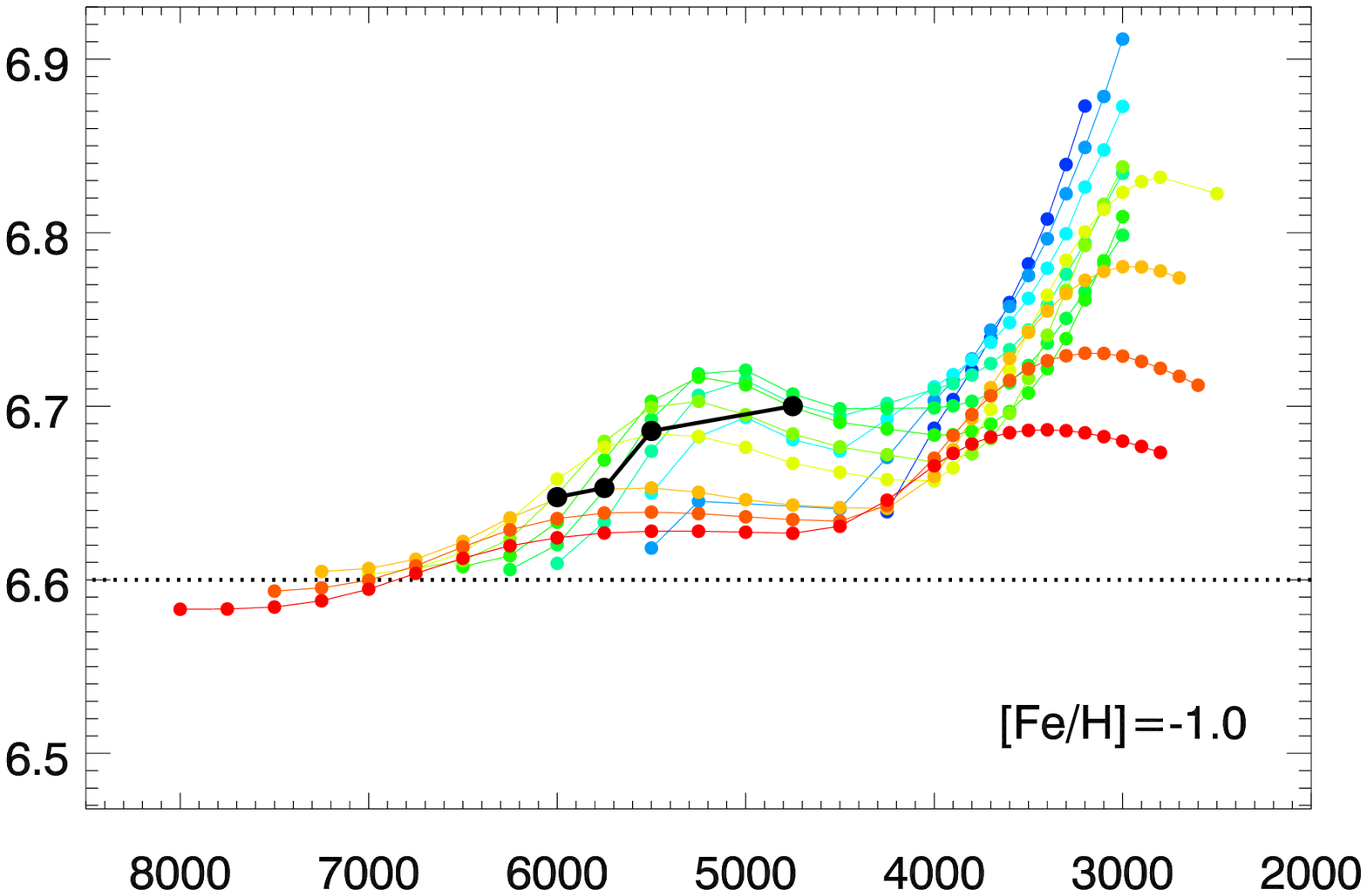}}
};
\node at (-0.29\textwidth,0.23\textwidth) {\scalebox{0.8}{5528 \AA}}; 
%\node [rotate=90] at (-0.35\textwidth,0.11\textwidth) {\scalebox{0.5}{A(Mg)$_{\rm{NLTE}}$}};
\end{tikzpicture}
&
\begin{tikzpicture}
\node[anchor=south east, inner sep=0] (image) at (0,0) {
\subfloat{\includegraphics[width=0.41\textwidth]{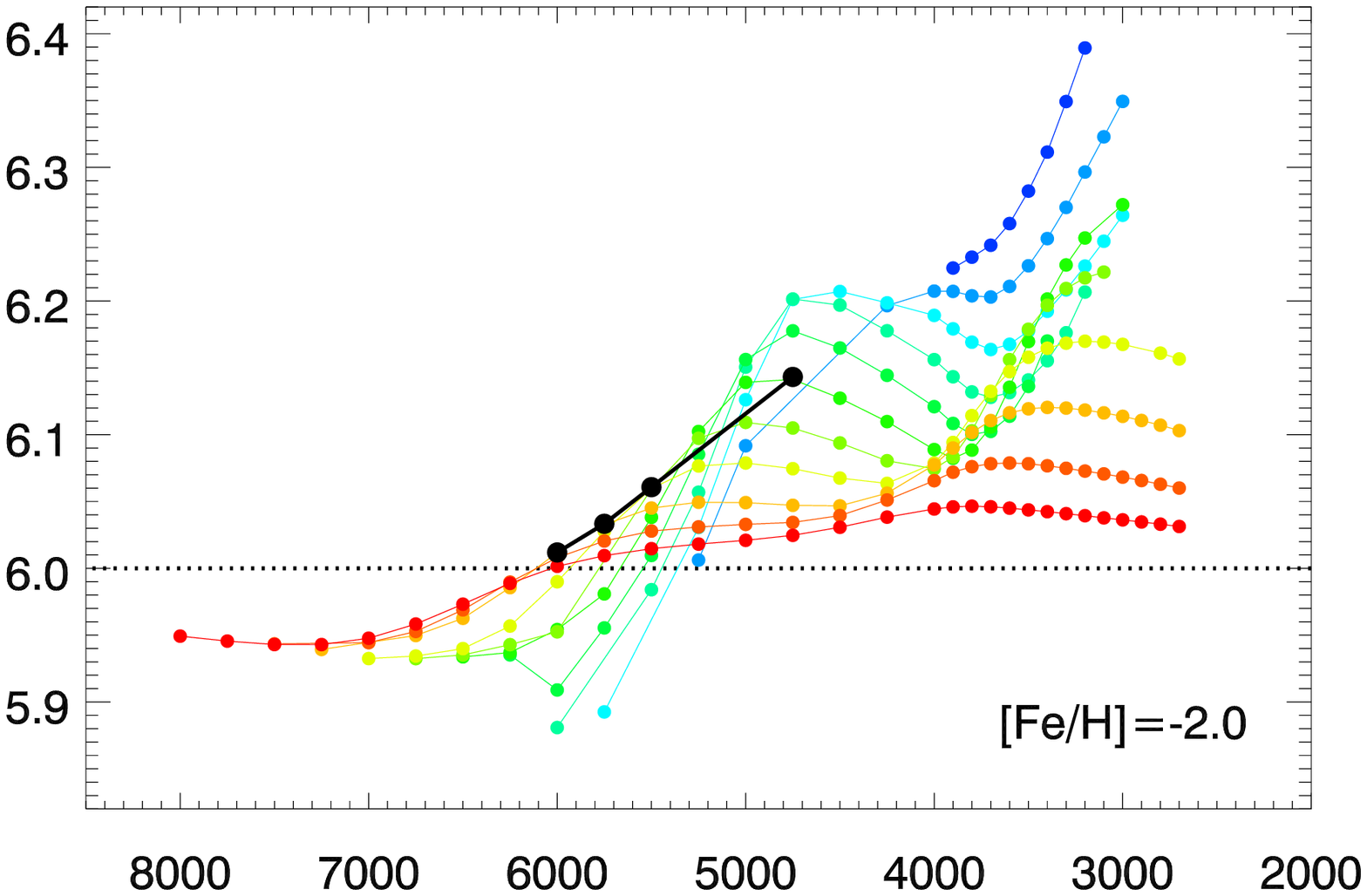}}
};
\node at (-0.29\textwidth,0.23\textwidth) {\scalebox{0.8}{5528 \AA}}; 
%\node [rotate=90] at (-0.35\textwidth,0.11\textwidth) {\scalebox{0.5}{A(Mg)$_{\rm{NLTE}}$}};
\end{tikzpicture} \\

\begin{tikzpicture}
\node[anchor=south east, inner sep=0] (image) at (0,0) {
\subfloat{\includegraphics[width=0.41\textwidth]{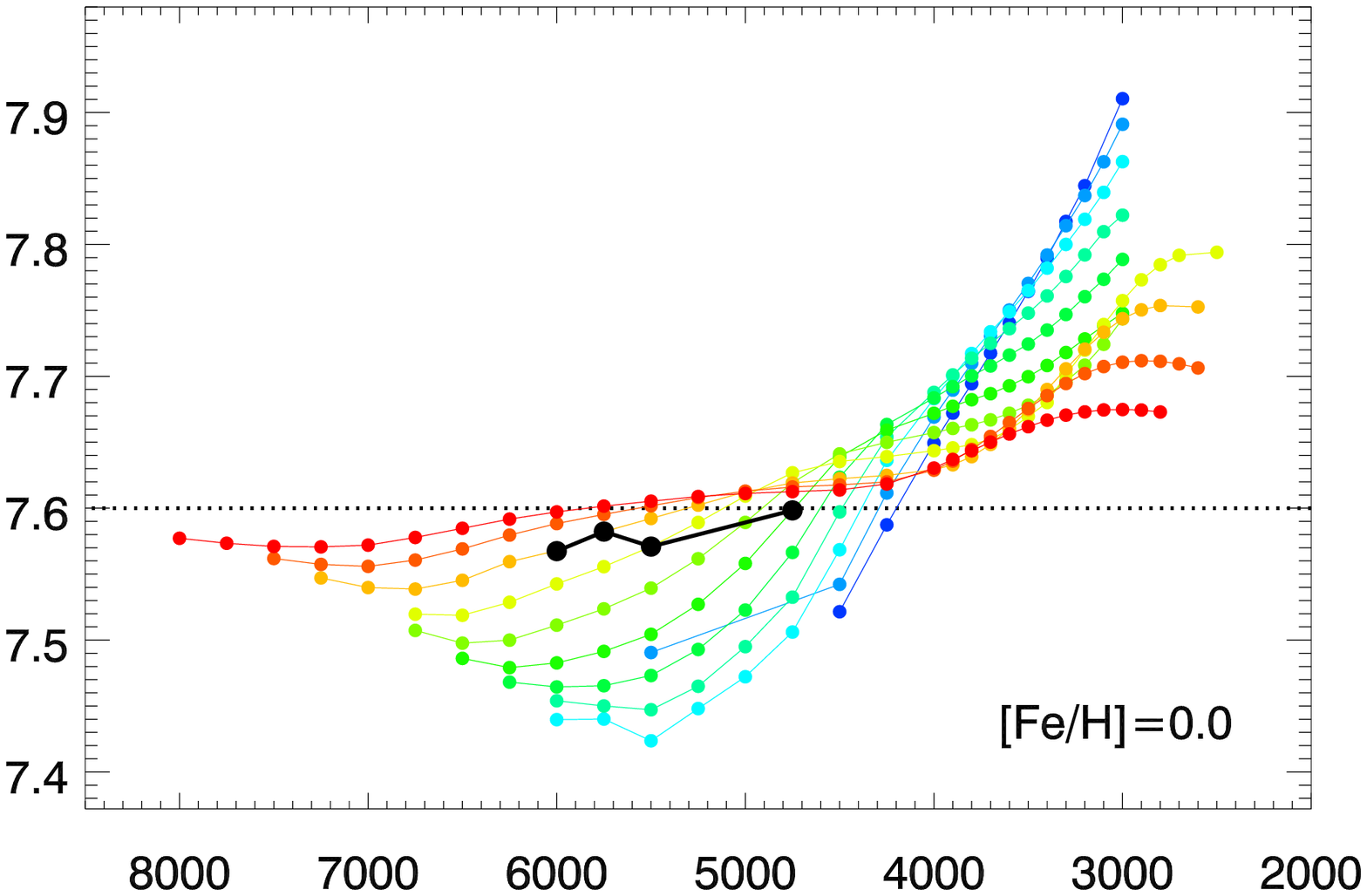}}
};
\node at (-0.29\textwidth,0.23\textwidth) {\scalebox{0.8}{5711 \AA}};
\node at (-0.15\textwidth,0.01\textwidth) {\scalebox{0.8}{\teff (K)}};  
\node [rotate=90] at (-0.36\textwidth,0.15\textwidth) {\scalebox{0.7}{A(Mg)$_{_{\rm{LTE}}}$}};
\end{tikzpicture}
&
\begin{tikzpicture}
\node[anchor=south east, inner sep=0] (image) at (0,0) {
\subfloat{\includegraphics[width=0.41\textwidth]{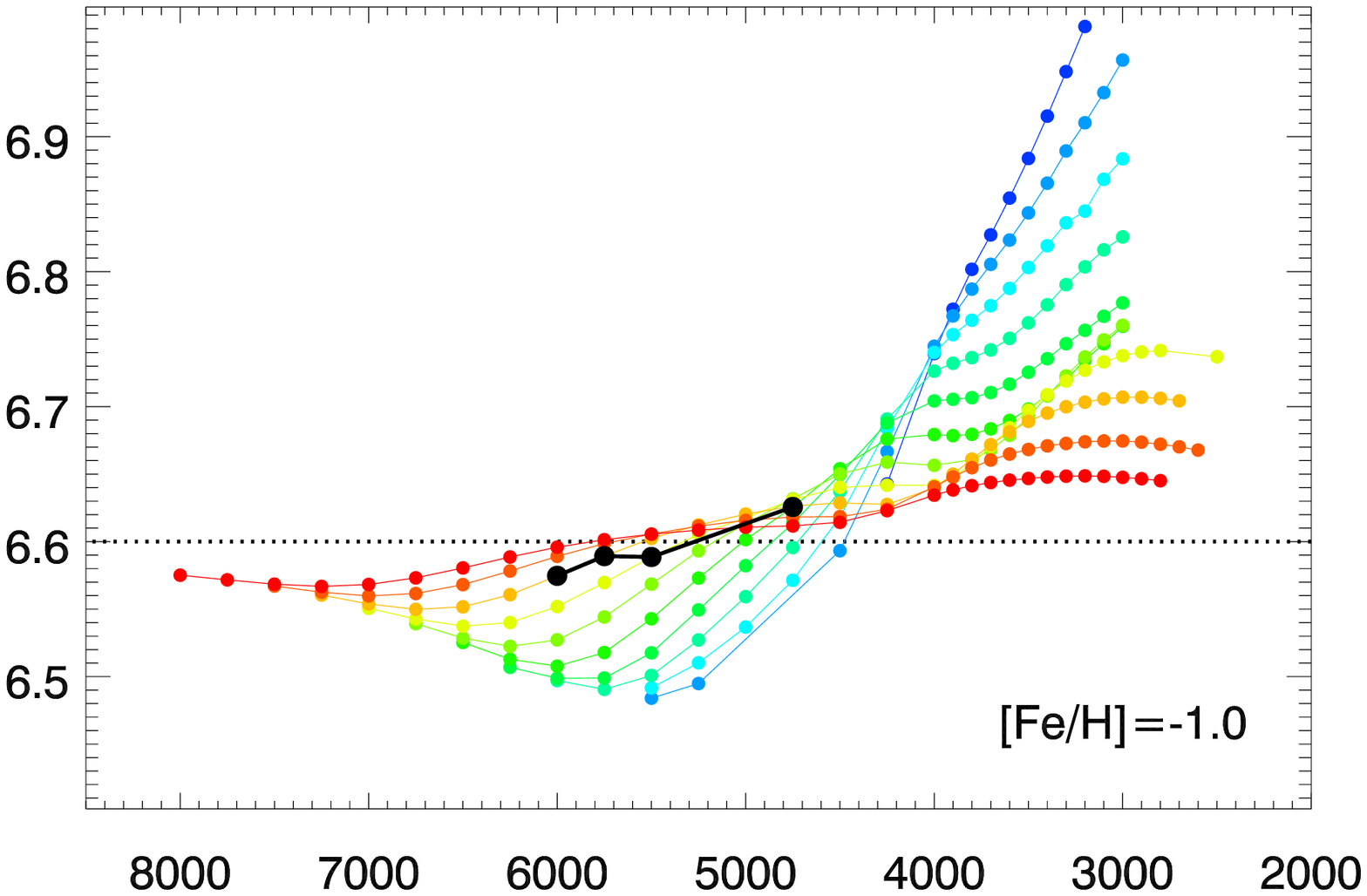}}
};
\node at (-0.29\textwidth,0.23\textwidth) {\scalebox{0.8}{5711 \AA}}; 
\node at (-0.11\textwidth,0.242\textwidth) {\scalebox{0.6}{\logg=0.5}};
\node at (-0.057\textwidth,0.238\textwidth) {\scalebox{0.6}{1.0}};
\node at (-0.059\textwidth,0.22\textwidth) {\scalebox{0.6}{1.5}};
\node at (-0.059\textwidth,0.20\textwidth) {\scalebox{0.6}{2.0}};
\node at (-0.059\textwidth,0.185\textwidth) {\scalebox{0.6}{2.5}};
\node at (-0.059\textwidth,0.175\textwidth) {\scalebox{0.6}{3.0}};
\node at (-0.043\textwidth,0.165\textwidth) {\scalebox{0.6}{3.5}};
\node at (-0.045\textwidth,0.152\textwidth) {\scalebox{0.6}{4.0}};
\node at (-0.044\textwidth,0.14\textwidth) {\scalebox{0.6}{4.5}};
\node at (-0.05\textwidth,0.13\textwidth) {\scalebox{0.6}{5.0}};
\node at (-0.15\textwidth,0.01\textwidth) {\scalebox{0.8}{\teff (K)}}; 
%\node [rotate=90] at (-0.35\textwidth,0.11\textwidth) {\scalebox{0.5}{A(Mg)$_{\rm{NLTE}}$}};
\end{tikzpicture}
&
\begin{tikzpicture}
\node[anchor=south east, inner sep=0] (image) at (0,0) {
\subfloat{\includegraphics[width=0.41\textwidth]{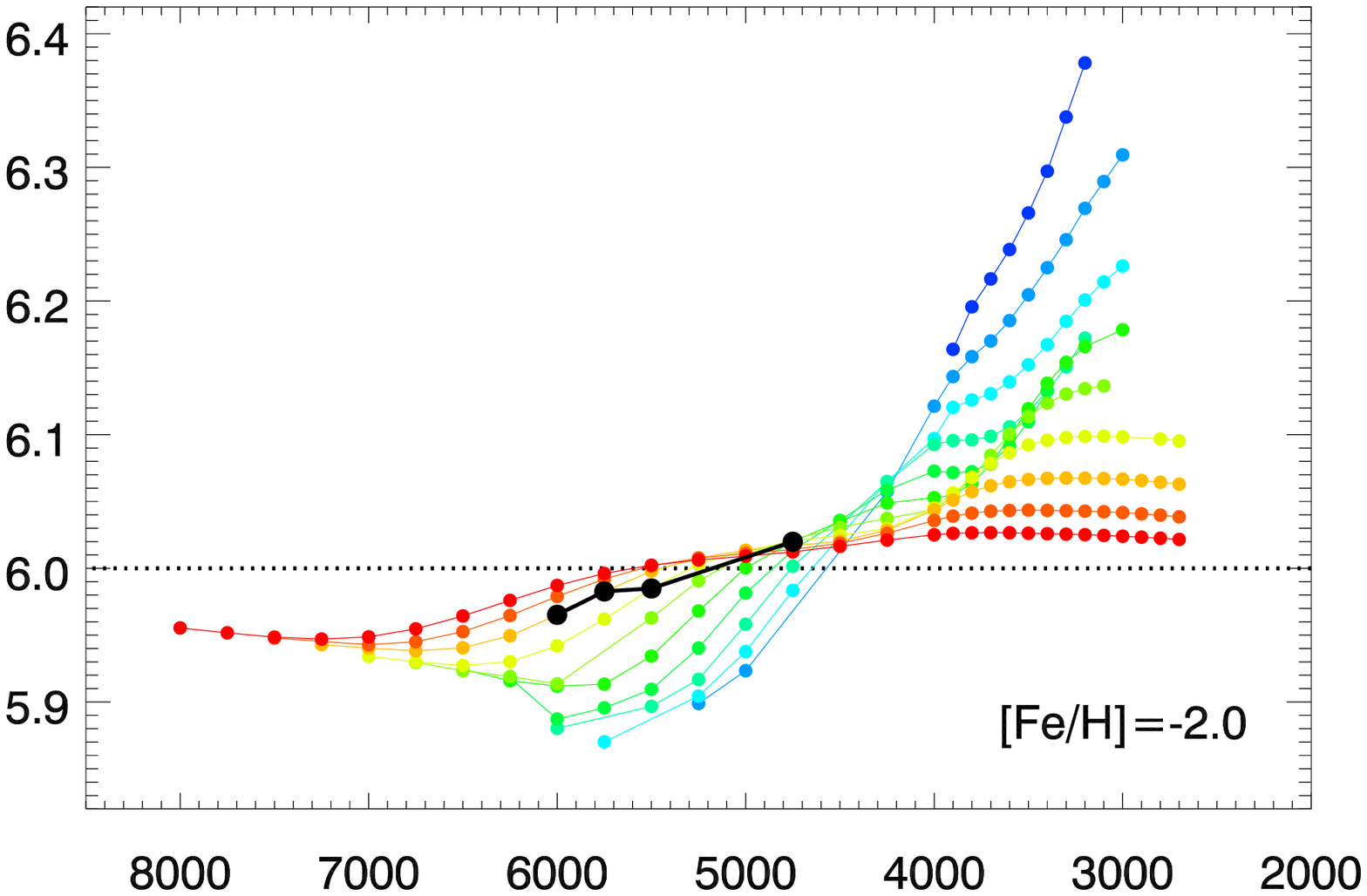}}
};
\node at (-0.29\textwidth,0.23\textwidth) {\scalebox{0.8}{5711 \AA}};
\node at (-0.15\textwidth,0.01\textwidth) {\scalebox{0.8}{\teff (K)}};  
%\node [rotate=90] at (-0.35\textwidth,0.11\textwidth) {\scalebox{0.5}{A(Mg)$_{\rm{NLTE}}$}};
\end{tikzpicture} \\
\end{tabular}
}
\caption{  For a fixed Mg non-LTE abundance (dotted line), the Mg abundance that would be derived in LTE is shown as function of \teff\ based on the 5528 (top) and 5711 (bottom) \AA\ lines. Colour lines connect model atmospheres with the same \logg(cm/s$^2$). \logg\ varies from 0.5 (blue) to 5.0 (red)~dex in steps of 0.5 dex.  The black points correspond to the RGB, bRGB, SGB and TOP models used in \fig{fig:diff}. }\label{fig:diffusion}
\end{figure*}

Comparison of the magnitude of evolutionary trends of Mg abundances seen in globular clusters \citep[e.g. Fig.~6 in][]{2007ApJ...671..402K} with the magnitudes of non-LTE effects seen, underlines the necessity to include accurate non-LTE corrections in order to study changes in atmospheric abundances with evolutionary status \citep[see also][]{2013A&amp;A...555A..31G,2014A&amp;A...567A..72G}. \fig{fig:diffusion} shows the Mg abundance one would derive in LTE as function of \teff\ for various values of \logg\ for a constant non-LTE abundance.  That is, it shows the predicted trend of abundance with evolutionary status that one would find  \emph{if}  non-LTE effects are neglected and \emph{if} the atmospheric Mg abundance were the same in the compared stars. 

\section{Conclusions}\label{sect:concl}

We obtained departure coefficients and equivalent widths for selected diagnostic Mg lines, which were presented and described.  The departure coefficient data are available on request from the authors, while the equivalent width data are presented in an electronic table on CDS. The non-LTE abundance corrections are usually small but not negligible. Giants tend to have negative corrections while dwarfs have in general positive corrections.  Abundance corrections increase at low metallicities.  These results are in general agreement with earlier studies; however, the magnitude of the corrections is often different (usually smaller) due to our improved collision data.  We have demonstrated the necessity to account for non-LTE for precise differential analysis between giants and dwarfs.

The non linearity of the non-LTE problem leads to a complicated error propagation and Table~\ref{tab:error} can be used as guide to uncertainties in non-LTE abundances due to collisional data in different regions of the stellar parameter space. In general uncertainties due to collisional data lead to uncertainties in derived abundances of $< 0.01$~dex  (2\%), although in some cases can be as large as 0.03 dex (7\%).  As these errors are less than or of the same order as typical corrections, we expect that we can use these results to extract Mg abundances from high quality spectra more reliably than from classical LTE analysis.

\begin{acknowledgements} 
The computations were performed on resources provided by SNIC through Uppsala Multidisciplinary Center for Advanced Computational Science (UPPMAX) under Project SNIC 2014/1-220. This work was supported by the Royal Swedish Academy of Sciences, the Wenner-Gren Foundation, G\"{o}ran Gustafssons Stiftelse and the Swedish Research Council. P.S.B. is a Royal Swedish Academy of Sciences Research Fellow supported by a grant from the Knut and Alice Wallenberg Foundation. P.S.B. was also supported by the project grant ``The New Milky'' from the Knut and Alice Wallenberg foundation.

\end{acknowledgements}

\bibliographystyle{aa}       		% style aa.bst	
\bibliography{MgNLTE,papers2}
\end{document}